\documentclass[sn-mathphys]{sn-jnl}

\usepackage{natbib}
\usepackage{siunitx}
\usepackage{graphicx}
\usepackage{xcolor}
\usepackage{graphicx} 
\usepackage{subfigure}
\usepackage{enumerate}

\DeclareFontFamily{U}{mathx}{\hyphenchar\font45}
\DeclareFontShape{U}{mathx}{m}{n}{<-> mathx10}{}
\DeclareSymbolFont{mathx}{U}{mathx}{m}{n}
\DeclareMathAccent{\widebar}{0}{mathx}{"73}

\jyear{2021}%

\theoremstyle{thmstyleone}%
\theoremstyle{thmstyletwo}%

\theoremstyle{thmstylethree}%

\begin{document}

\title[Learning an Adaptive Forwarding Strategy for Mobile Wireless Networks]{Learning an Adaptive Forwarding Strategy for Mobile Wireless Networks:  Resource Usage vs. Latency}


\author*[1]{\fnm{Victoria} \sur{Manfredi}}\email{vumanfredi@wesleyan.edu}

\author*[1]{\fnm{Alicia P.} \sur{Wolfe}}\email{pwolfe@wesleyan.edu}

\author[2]{\fnm{Xiaolan} \sur{Zhang}}\email{xzhang@fordham.edu}

\author[3]{\fnm{Bing} \sur{Wang}}\email{bing@uconn.edu}

\affil*[1]{\orgdiv{Department of Mathematics and Computer Science}, \orgname{Wesleyan University}, \orgaddress{\street{265 Church Street}, \city{Middletown}, \postcode{06459}, \state{Connecticut}, \country{USA}}}

\affil[2]{\orgdiv{Department of Computer and Information Sciences}, \orgname{Fordham University}, \orgaddress{\street{441 E. Fordham Road}, \city{Bronx}, \postcode{10458}, \state{New York}, \country{USA}}}

\affil[3]{\orgdiv{Computer Science \& Engineering Department}, \orgname{University of Connecticut}, \orgaddress{\street{371 Fairfield Way}, \city{Storrs}, \postcode{06269}, \state{Connecticut}, \country{USA}}}

\abstract{
Mobile wireless networks present several challenges for any learning system, due to uncertain and variable device movement, a decentralized network architecture, and  constraints on network resources. In this work, we use deep reinforcement learning (DRL) to learn a scalable and generalizable forwarding strategy for such networks. We make the following contributions: i) we use hierarchical RL to design DRL packet agents rather than device agents to capture the packet forwarding decisions that are made over time and improve training efficiency; ii) we use relational features to ensure generalizability of the learned forwarding strategy to a wide range of network dynamics and enable offline training; and iii) we incorporate both forwarding goals and  network resource considerations into packet decision-making by designing a weighted reward function. Our results show that the forwarding strategy used by our DRL packet agent often achieves a similar delay per packet delivered as the oracle forwarding strategy and outperforms all other strategies (including state-of-the-art strategies) in terms of delay, even on scenarios on which the DRL agent was not trained.}



\maketitle

\section{Introduction}
\label{sec:introduction}

Mobile wireless networks have been used for a wide range of real-world applications, from vehicular safety \cite{Toh01:adhoc,Sommer14:vehicular,Gerla14:vehicular} to 
animal tracking \cite{Juang02:zebranet,Zhang04:zebranet} to environment monitoring \cite{Oliveira2021:forest,Albaladejo2010:ocean} to search-and-rescue \cite{Huang05:CenWits,Jiang2009:SenSearch,robinson2013resilient} to military deployments \cite{ramanathan2010scalability,poularakis2019flexible} to the mobile Internet of Things \cite{bello2014intelligent}. These networks present several challenges, however,  for  learning systems. 
Devices are moving due to their association with, for instance,  vehicles, 
robots, animals,  
or  people, which causes changes in the network connectivity.
As a consequence, devices are often only able to communicate 
with each other during limited windows of time when devices are within transmission range of each other. Furthermore, it may be difficult to predict when  opportunities for communication may occur as these depend on the dynamics of device movement.
Depending on the specific network problem to address, there may also be competing goals to trade-off, such as minimizing packet delay vs. device resource usage.
Finally, the network architecture is decentralized,  complicating sharing of network state (and thus training of learning systems) as exchange of  information is limited by the communication opportunities available in the mobile wireless network.



\begin{figure*}[t]
    \centerline{\hbox{
        \subfigure[{\em Routing when devices are stationary.} Because devices are stationary, once end-end paths are established, they very rarely need to be updated. Here, $v$ has a contemporaneous end-end path to  $d$, and so packet $p$ is forwarded along the path.]
        {\includegraphics[width=4.7in, trim = 0.2cm 0.3cm 7.8cm 14.cm, clip]{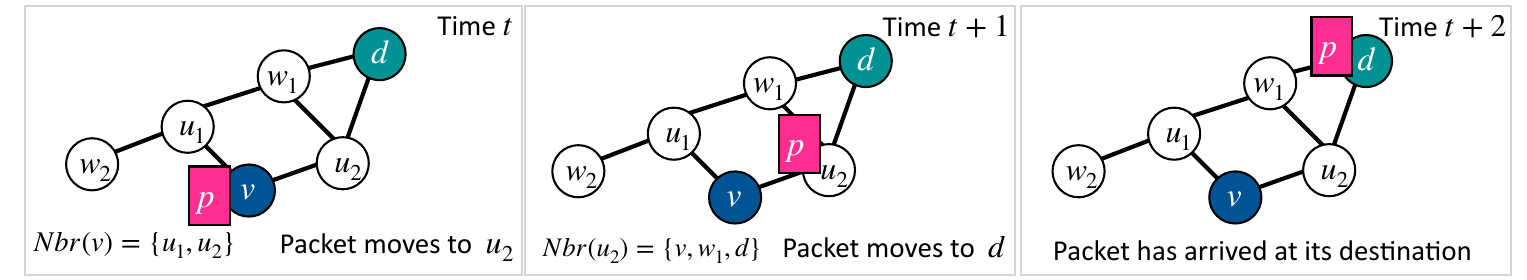}}
    }}
 \centerline{\hbox{
        \subfigure[{\em Routing when devices are mobile and the network is well-connected}. Despite mobility, network connectivity is sufficiently stable that end-end paths can still be established and used for forwarding traffic, though may need to be periodically re-established. Here again, $v$ has a contemporaneous end-end path to  $d$, and so packet $p$ is forwarded along the path, though $v$'s path to $d$ may be different in the future.]
        {\includegraphics[width=4.7in, trim = 0.2cm 13.9cm 7.8cm 0.5cm, clip]{figs/mobile}}
    }}    
    \centerline{\hbox{
        \subfigure[{\em DTN forwarding when devices are mobile and the network is poorly connected}.  
        Now contemporaneous end-end paths rarely exist. Only temporal paths exist from  $v$ to $d$, so devices independently choose packet $p$'s next hop and $p$ is forwarded  hop-by-hop to  $d$.]
        {\includegraphics[width=4.7in, trim = 0.2cm 6.9cm 7.8cm 7.4cm, clip]{figs/mobile}}
    }}
\caption{Packet $p$ destined to device $d$  travels from device $v$ to device $u_2$ to finally arrive at its destination device $d$, differences in stationary and mobile wireless networks (see (a) vs. (b) and (c)).  In mobile wireless networks (i.e., (b) and (c)), whether routing or DTN forwarding is used depends on whether a contemporaneous path is present.}
\label{fig:mobile-ex}
\end{figure*}

In this work, we   focus on the problem of packet forwarding in a mobile wireless network.  Traditionally, forwarding strategies are hand-crafted to  target specific kinds of  network connectivity. 
In static networks,  see Fig. \ref{fig:mobile-ex}(a), once  end-end paths are discovered, these paths are generally stable as the network connectivity does not change.
In comparison, while end-end paths may occasionally exist  in some mobile networks due to dense connectivity or slow device movement, see Fig. \ref{fig:mobile-ex}(b),  these paths  typically only exist for short periods of time and are periodically re-established using  ad hoc routing algorithms \cite{Johnson96dynamicsource,perkins2003rfc3561,clausen2003optimized,Perkins94:DSDV}. 
In other mobile networks, devices  meet only occasionally due to sparse connectivity or fast device movement, see Fig. \ref{fig:mobile-ex}(c), and so contemporaneous end-end paths rarely exist. Consequently, delay tolerant network (DTN) forwarding algorithms \cite{singlecopy-spyro, multicopy-spyro, vahdat20:epidemic} are used to select the best next hop for a packet using criteria such as expected delay to meet the packet's destination.

In many real mobile wireless networks, however, a mix  of connectivity is often found, see \cite{manfredi2011understanding}, motivating the need for an adaptive forwarding strategy.
Forwarding strategies that do explicitly adapt to disparate network conditions often focus on switching between two different strategies, such as between  ad hoc routing and flooding \cite{danilov2012adaptive, seetharam2015routing},  or between ad hoc routing and delay tolerant forwarding  \cite{lakkakorpi2010adaptive, delosieres2012batman, raffelsberger2014combined, asadpour2016route}. Strategy switching, however, risks instability and poor convergence  if conditions change quickly or  network state is only partially observable. 

To address the above challenges,  we use  deep reinforcement learning  (DRL) \cite{sutton2018reinforcement} to design an adaptive packet forwarding strategy, using deep neural networks (DNNs) \cite{bengio2009learning} to approximate the RL policy. 
By choosing next hops locally at devices, both contemporaneous  and temporal paths can be implicitly constructed by the DRL agent. Importantly,  devices located in different parts of a mobile network 
can each  independently run the same DRL agent as the agent will  react appropriately  to the local state at each device by using the parts of its policy relevant for that state. Consequently, a single policy can be applied to a state space that includes both well-connected and poorly connected parts of a mobile network.

Most works on DRL-based forwarding in wireless networks focus on  stationary devices; those  that do consider device mobility either target specific kinds of network connectivity or have other limitations (see Section \ref{sec:related-work}). In this work, we specifically
focus on designing a  forwarding strategy for mobile wireless networks able to adapt to very different kinds of network connectivity. 
We make the following contributions.



\vspace{0.1in}
\begin{enumerate}[i)]

\item {\em Packet-centric decision-making to simplify learning (Section \ref{sec:deeprl}).} 
We use {\em packet agents}, making packets, rather than devices, the  decision-making agent,  to accurately assign credit to those actions that affect packet success. Because packets can wait for several time steps before making a decision,  we use hierarchical RL \cite{dietterich1998maxq, sutton1999between,barto2003recent} with {\em fixed policy options} \cite{sutton1999between} to more efficiently back up from one decision point to another.


\item{\em Relational features to ensure generalizability despite device mobility (Section \ref{sec:mobile-features}).}
We use   {\em relational features} to represent the states and actions used by the DRL agent. Relational features model the relationship between network devices instead of describing a specific device, and so support generalizability to scenarios on which the DRL agent was not trained.
This ability to generalize also allows us to {\em train offline} thereby avoiding the need for decentralized communication. Relational features also allow us to structure the DNN representing the DRL forwarding policy to consider {\em one action at a time}, producing a single Q-value per state and action pair. The number of times the DNN is used to predict Q-values then corresponds to the number of actions available, allowing the trained DRL agent to handle varying numbers of neighbors (see Fig. \ref{fig:mobile-ex}) and so be network independent.


\item {\em A weighted reward function to trade-off competing goals (Section \ref{sec:mdp}). } 
To incorporate packet forwarding goals and network resource considerations 
into packet decision-making,  we design a {\em weighted reward function} for the DRL agent. Using our reward function, more weight can be placed on higher priority forwarding considerations, such as to  not waste network resources unnecessarily. 

\item{\em Extensive evaluation (Section \ref{sec:simulation}).}
We evaluate our approach, which combines the above key design decisions, 
on two widely-used mobility models with varying numbers of devices and transmission ranges,
spanning the continuum from disconnected to well-connected networks. 
Our results show that our DRL agent trained on only one of these scenarios generalizes well to the other scenarios including for another mobility model.
Specifically, our DRL agent often achieves delay similar to an oracle strategy and almost always outperforms all other strategies in terms of delay including the state-of-the-art seek-and-focus strategy \cite{singlecopy-spyro}, even on scenarios on which the DRL agent was not trained.  While the oracle strategy requires global knowledge of future device movement, our DRL agent uses only locally obtained feature information.

\end{enumerate}
\vspace{0.1in}




The rest of this paper is organized as follows. 
In Section \ref{sec:related-work}, we overview related work on DRL-based forwarding.
In Section \ref{sec:rlagent}, we describe how we formulate a DRL model able to learn an adaptive forwarding strategy.
In Section \ref{sec:simulation}, we present simulation results using our DRL model.
Finally, in Section \ref{sec:conclusions}, we summarize our main results and provide directions for future research.

\section{Related Work}
\label{sec:related-work}

In this section, we overview related work on routing and forwarding strategies for mobile wireless networks.  We divide these strategies into two classes: {\em hand-crafted} strategies not based on learning, and {\em RL-based} strategies which use RL to make decisions.  We additionally focus on single-copy strategies, to allow operation in not just sparsely but also densely connected networks.
In contrast, multi-copy strategies, in which
 multiple copies of the same packet may be made to reduce packet delivery delay, typically focus on only very sparse network scenarios as the risk of introducing congestion from making packet copies is lower than in dense networks. 

\subsection{Hand-crafted Approaches}

Routing strategies like DSR~\cite{Johnson96dynamicsource}, AODV~\cite{perkins2003rfc3561}, OLSR~\cite{clausen2003optimized} and DSDV~\cite{Perkins94:DSDV} have been developed for ad hoc networks (see Fig. 
\ref{fig:mobile-ex}(b)). Ad hoc  networks are typically  well-connected so that at least one 
contemporaneous  end-end path exists between a source and  destination. In delay tolerant networks~\cite{Jain2004:DTN}, aka opportunistic networks, the network topology is  so sparse that no contemporaneous
path exists between a source and destination and so researchers have developed epidemic flooding~\cite{vahdat20:epidemic} as well as more resource-efficient forwarding schemes like the Seek-and-Focus and Utility-Based strategies~\cite{singlecopy-spyro}. 
All of these strategies, however, are primarily designed for networks with certain conditions (e.g., persistent or non-persistent network connectivity), and require a priori knowledge of the network conditions of the network before being chosen.
In our simulation results in Section \ref{sec:simulation}, we compare our approach  with the  Seek-and-Focus and Utility strategies of \cite{singlecopy-spyro} as these are state-of-the-art for single-copy routing in delay tolerant networks. We also compare  with an oracle strategy that gives the best possible performance though is not practicably implemented.

Forwarding strategies that do explicitly adapt to disparate network conditions have been less explored and often focus on switching between two different strategies, such as between  ad hoc routing and flooding \cite{danilov2012adaptive, seetharam2015routing},  or between ad hoc routing and delay tolerant forwarding  \cite{lakkakorpi2010adaptive, delosieres2012batman, raffelsberger2014combined, asadpour2016route}. Strategy switching, however, 
can lead to instability and poor convergence  if conditions change quickly or  network state is only partially observable.

A number of works focus on designing multi-copy forwarding strategies, for instance \cite{multicopy-spyro, tie2011r3, yang2016hybrid}. Multi-copy strategies generate multiple copies of the same packet to increase the probability that at least one copy of the packet is delivered within a certain amount of time. While creating packet copies  improves delivery delay for a given packet, doing so also increases the traffic load on the network, and hence congestion.  Consequently, multi-copy strategies typically focus on very sparse network scenarios in which devices only occasionally have neighbors and there is little traffic, making the risk of congestion from packet copies lower.
Because our focus is on designing forwarding strategies that seamlessly adapt between very sparse and very dense network connectivity with possibly significant amounts of traffic, we focus in this work on designing a single-copy forwarding strategy (i.e., no packet copies are made).

\subsection{DRL-Based Approaches}

DRL-based approaches to routing and forwarding in mobile and wireless networks 
have the advantage of making it easy to take into consideration different network features and optimization goals that may be difficult for humans to reason about. DNNs specifically provide a natural way to unify different approaches to modeling mobility for forwarding decisions: features that work well for one kind of mobility can easily be included along  with features that work well for other kinds of mobility.  The DNN representation itself supports generalization to unseen types of device mobility and network scenarios. Through training, a DRL agent learns the relationship between mobile network features and how best  to route, and encodes that in a policy represented using a DNN. 

Early works  \cite{Boyan94:Q-routing,Choi96:PQ-routing,Kumar98:Confidence,hu2010adaptive,elwhishi2010arbr,rolla2013reinforcement}   focus on distributed routing
but use less scalable and less generalizable table-lookup based approaches and generally learn online. When the network topology is changing, online learning can be problematic as now devices may have few or no neighbors and there may be limited bandwidth to exchange any information for training.
Consequently, many of the recent works on DRL-based  forwarding strategies focus on  stationary networks where devices do not move, see \cite{ye2015multi,Valadarsky17:learning,Xu2018:experience, di2019carma,Mukhutdinov19:Multi-agent, Suarez-Varela19:feature,you2019toward,you2020toward,chen2021multiagent,wowmom2021-relational,almasan2022deep}.

Fewer works consider forwarding in 
mobile wireless networks, and those that do often optimize for specific kinds of network connectivity, such as focusing primarily on vehicular networks \cite{li2018hierarchical,schuler2021robust,lolai2022reinforcement,luo2021intersection} or UAV networks \cite{feng2018multi,sliwa2021parrot,rovira2021fully,schuler2021towards,qiu2022data}. 
Works that consider mobile networks more broadly have limitations:
\cite{johnston2018reinforcement} 
extends early work on strategies that learn online \cite{Boyan94:Q-routing,Kumar98:Confidence}
to tactical network environments but uses RL to estimate the shortest path to the destination and focuses on multi-copy forwarding, i.e., making additional copies of packets  to reduce delay, unlike the single-copy forwarding strategy we design in this work;
\cite{sharma2020rlproph} focuses on sparse network scenarios,
specifically delay tolerant networks;
\cite{han2021qmix}  focuses on forwarding messages to network communities rather than individual devices in delay tolerant networks; 
\cite{jianmin2020ardeep,kaviani2021robust}  focus on relatively limited network scenarios with a few fixed flows and up to 50 devices.

\smallskip
\noindent{\bf {Differences from prior work.}} The goal of our work 
is to design an adaptive DRL-based forwarding strategy that can span the continuum from sparsely connected to well-connected mobile networks. 
In Section \ref{sec:simulation}, we show that our learned forwarding strategy trained on a  mobile wireless network with 25  devices can generalize during testing to
mobile networks  with 100 devices and varying transmission ranges. While our work builds off of ideas in \cite{wowmom2021-relational}, we consider the much harder network setting of mobile devices rather than the stationary devices considered in \cite{wowmom2021-relational}, and we design novel  features to capture temporal and spatial network connectivity as well as propose a reward function  to reflect competing network goals.
Our use of offline centralized training of the DRL agent 
(but online distributed operation) 
mitigates online training's need to exchange significant amounts of information (such as whether a packet was delivered or dropped) between devices separated by many hops. When deployed in a network, each device independently uses its own copy of the trained RL agent to make distributed routing decisions.

    

An important takeaway of our work is the need to tailor  existing machine learning techniques to handle the decentralized communication and resource constraints of mobile wireless networks.
For instance,  while we have some features that represent actual relationships between devices as in relational learning \cite{getoor2007introduction,koller2007introduction,struyf2010relational}, we primarily use relational techniques to treat devices as interchangeable objects described by their attributes rather than their identities, to build a single model that works across all devices.
While our aggregated neighborhood features are similar in spirit to graph neural networks  \cite{scarselli2008graph,Thomas2017iclr,google-gnn}, our features are simpler to compute 
and can more easily handle changing neighbors. 
While we use DRL to learn a policy, we handle actions differently than in a DQN \cite{mnih2015human}
since the number of actions available at a device changes over time and varies across devices.
Finally, while our weighted reward function could also be converted to use multi-objective reward techniques \cite{van2014multi,hayes2022practical},  which find policies for a range of reward factor weightings, 
doing so would make training the DRL agent more computationally expensive.
\section{Learning an Adaptive Forwarding Strategy}
\label{sec:rlagent}

In this section, we overview our approach to using DRL to learn an adaptive forwarding strategy. 
We first describe how we formulate the problem of forwarding packets  as a learning problem (Section \ref{sec:deeprl}). 
Then, we describe the relational features we use to model a mobile wireless network (Section \ref{sec:mobile-features}).
Next, we describe how 
we construct the DRL agent's states and actions from the relational features, and derive a reward function for packet forwarding (Section \ref{sec:mdp}).
We finish by describing how our DRL agent makes forwarding decisions (Section \ref{sec:decision-making}) and how offline training of the DRL agent is performed (Section \ref{sec:offline-training}).


\subsection{Packet Forwarding Using DRL}
\label{sec:deeprl}

RL focuses on the design of intelligent agents:  an RL  agent interacts with its environment to learn a policy, i.e., which actions to take in different environmental states. The environment is modeled using a Markov decision process (MDP). An MDP comprises a set of states ($S$), a per state set of actions ($\mathcal{A}(s)$), a reward function, and a Markovian state transition function in which the probability of the next state $s' \in S$ depends only on the current state $s \in S$ and action $a \in \mathcal{A}(s)$.  RL assumes that these state transition probabilities are  not known, but that samples of transitions of the form $(s, a, r, s')$ can be generated.
From these samples, the RL algorithm learns a $Q$-value  for each $(s, a)$ pair. 
A $Q$-value estimates the expected future reward for an RL agent, when starting in state $s$ and taking action $a$. Once learned, the optimal action in state $s$ is the one with the  highest $Q$-value.  

\begin{figure*}[t]\centering 
   \includegraphics[width=4.7in, trim = 0.45cm 7.cm 5.5cm 5.cm, clip]{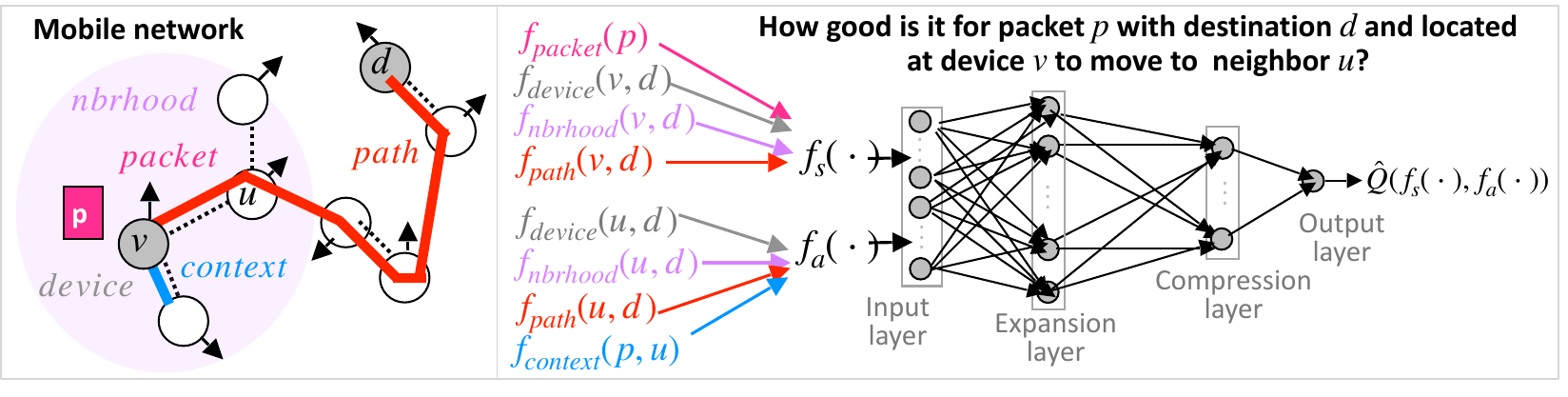}
\caption{Overview of how our DRL packet agent makes decisions. Packet $p$  at device $v$ makes a forwarding decision by  activating the DNN at $v$ once for each action $a$ available in $p$'s current state $s$. The DNN inputs are the features $f_s(\cdot)$ and $f_a(\cdot)$, which describe the state from $p$'s point of view and the action under consideration. Packet  $p$ chooses the action with the best estimated Q-value. }
\label{fig:routing-architecture}
\end{figure*}

When the MDP has a small number of states and actions, an RL agent can learn a $Q$-value function using Q-learning (see \cite{watkins1992q}). When the state space is too large for exact computation of the $Q$-values, function approximation may be used to find approximate $Q$-values. Here, we use DNNs  for function approximation, see Fig. \ref{fig:routing-architecture}, as in Deep RL (DRL). Each state $s$ and action $a$ is translated into a set of features via the functions $f_s(\cdot)$ and $f_a(\cdot)$. These features are 
input to the DNN, to produce as output an approximate $Q$-function $\hat Q(f_s(\cdot), f_a(\cdot))$.

In a mobile wireless network, it is natural to think of devices as the DRL agents choosing next hops for packets, but doing so confuses the decision-making dependencies that are involved in forwarding a packet.
That is, the steps from a packet's source device to its destination device (or drop) are constructed by the behavior of all devices through which the packet passes, rather than only by the  device at which the packet is currently located.
Thus, we use {\em packet agents}, making packets, rather than devices, the  decision-making agent,  to more accurately assign credit to the actions that affect packet success. Devices now only choose which packets should have the opportunity to make a decision, but do not choose the next hop for a packet.

Packet-centric decision-making can be viewed as a multi-agent problem, as each packet interacts with others while attempting to greedily optimize its own travel time. We do not use a global cooperative reward function, however. Instead, we have each device queue enforce fairness among its packets, choosing which packet gets to make a decision about its next hop. In this work, devices allow the first $k$ packets to make a decision, where $k$ is generally sufficiently large that all packets are forwarded. Alternative queuing disciplines include allowing only the packet at the front of the queue to go or selecting a subset of packets based on the Q-value for each packet's best action combined with a fairness metric quantifying the number of decisions made by packets in the same flow (i.e., going from the same source to the same destination).

A packet's decision-making, however, can also 
involve multiple time steps, such as waiting in a queue for its turn to make a forwarding decision, or waiting for another device to come within transmission range. Thus, we use hierarchical RL \cite{dietterich1998maxq, sutton1999between,barto2003recent}, specifically {\em fixed policy} options \cite{sutton1999between}, where the same single action is the only action available at every time step for the duration of the option.
Our use of fixed policy options allows reward signals to be backed up over multiple time steps in a single update. The use of options additionally 
supports faster learning while using less training data.

\subsection{Mobile Network Features}
\label{sec:mobile-features}

Our goal is to be able to use the same learned forwarding strategy  at different devices and in different mobile networks (e.g., dense or sparse networks with or without contemporaneous end-to-end paths) with unknown or unseen device mobility.  To achieve this, we use {\em relational features} to represent the states and actions used by the DRL agent.  Relational features, such as delay to destination, model the relationship between network devices instead of describing a specific device and so are not tied to a specific network topology. 

We propose five classes of relational features specifically tailored to model mobile wireless networks. These classes give a general framework for organizing the features needed for forwarding packets in a mobile network.  The features we propose for each class, however, are not exhaustive but rather only the specific features that our DRL agent uses in the simulation results in Section \ref{sec:simulation}; we expect many other features could be proposed for each class. 
We next describe the features we use, from the point of view of a  packet $p$ when choosing its next hop. We assume that $p$'s 
destination is device $d$ and that $p$ is currently located at device $v$ which has neighbors $u \in Nbr(v)$. 

\vspace{0.1in}
\begin{enumerate}
\item {\em Packet features, ${f}_{packet}(p)$}, are  a function of information about packet $p$. We propose the following packet features. 
\vspace{0.05in}

\begin{enumerate}[(i)]
\item Packet $p$'s {\em time-to-live (TTL)}. TTL is a packet header field used  in real networks to prevent the possibility of 
packets looping  forever in the network. 
When a packet is generated at its source device, 
the packet's TTL field is set to some initial value. Then, the  TTL field is decremented by one whenever a packet is forwarded to another device. 
A packet is dropped when its TTL reaches 0.
\item Packet $p$'s {\em time-at-device}, that is how long $p$ has been at its current device, $v$.
\end{enumerate}

\vspace{0.1in}
\item {\em Device features, ${f}_{device}(v, d)$}, are a function of device $v$'s information  and  destination device $d$'s ID. We propose the following device features computed at the current time step $t$.
\vspace{0.05in}

\begin{enumerate}[(i)]
\item Device $v$'s {\em queue length}. Even when a network has little congestion,  queue length information can still be useful when choosing next hops.  For instance, next hop devices that do not have any packets (or any packets in the same traffic flow as $p$) may be preferred  to better distribute traffic and decrease delivery delay.

\item Device $v$'s {\em per-destination queue length} considering only packets for destination $d$.
\item Device $v$'s  {\em node degree}, corresponding to the number of neighbors of device $v$,  that is, the number of devices within wireless communication range of $v$.

\item Device $v$'s  {\em node density}, computed as the fraction of neighbors that $v$ has out of the $N$ devices in the network.

\end{enumerate}
\vspace{0.05in}
Our DRL agent additionally keeps track of two device features, the $x$ and $y$ location coordinates
of device $v$,  which are not input into the DNN but are only used to compute the Euclidean distance, a path feature described later in this section. 
Even when a device  knows its own location,  locations for  other devices can only be obtained when two devices meet and exchange features. 
Consequently, location (and thus distance) features can be out-of-date. 




\vspace{0.1in}
\item{\em Path features, $f_{path}(v, d)$}, describe the time-varying path from a device $v$ to a destination device $d$. Unlike the other features discussed so far, path features can use not just current device information but also historical information.
 Consequently, these features may have some associated uncertainty.
We propose the following path features. 
 \vspace{0.05in}
\begin{enumerate}[(i)]
\item {\em Last inter-meeting time} between device $v$ and destination $d$.
\item {\em Last meeting duration} between device $v$ and destination $d$.
\item {\em Euclidean distance} from device  $v$ to  destination $d$.  
Euclidean distance is calculated  using $v$'s current $x$ and $y$ location coordinates, and device $v$'s recorded (and possibly out-of-date) location coordinates of destination $d$ (see description of device features).

\item  {\em Timer transitivity} between device $v$ and destination $d$, as defined in \cite{singlecopy-spyro} for utility-based and seek-and-focus forwarding. 
Each device $v$ maintains a timer for each other device $d$ in the network, denoted as $\tau_v(d)$, which is the time elapsed since device $v$ last met device $d$.
We implemented the timer transitivity as defined in \cite{singlecopy-spyro}: when two devices, $u$ and $v$ encounter each other, if $\tau_u(d) < \tau_v(d) - t(d_{u,v})$, where $t(d_{u,v})$ is the expected time for a device to move a distance of $d_{u,v}$ (the distance between devices $u$ and $v$), then $\tau_v(d)$ is set to  
$\tau_v(d) = \tau_u(d) + t(d_{u,v})$.
When the device locations are not known, the distance  $d_{u,v}$ can be approximated using the transmission range: device $v$ can only be so far away from device $u$ for $v$ to be within the transmission range of $u$.
The insight is that  for many mobility models, a smaller timer value on average implies a smaller distance to the device, where the timer  evaluates the ``utility'' of a device in delivering a packet to another device. 
In experiments we have done, we have observed correlation between timer values and distance. Consequently, timer transitivity  can be used as a feature to approximate distance when location coordinates, and consequently the Euclidean distance feature, are not available.

\end{enumerate}

\vspace{0.1in}
\item{\em Neighborhood features, ${f}_{nbrhood}(v, d)$}, are computed over the current neighbors  $Nbr(v)$ of a device $v$.
We propose the following  neighborhood features:
for each device and path feature, $f_{i} \in f_{device}(v, d) \cup f_{path} (v,d)$, we compute the minimum, maximum, and average  over device $v$'s current neighborhood, $Nbr(v)$.
This is similar in spirit to the aggregation function in a 
graph neural network \cite{scarselli2008graph,Thomas2017iclr,google-gnn}.
These features compress the information obtained from a variable number of neighbors into a fixed size vector  to  input to the  DNN in Fig. \ref{fig:routing-architecture}.

\vspace{0.1in}
\item{\em Context features, ${f}_{context}(p, u)$},  provide context for other features. For a packet $p$ at a device $v$ considering a  next hop $u \in Nbr(v) \cup v$, we propose context features that indicate whether $p$ has recently visited $u$ or not.
    Each packet $p$ stores in its packet header the last $N_{history}$ device IDs that it visited,  where $N_{history}$ is a predetermined constant.
    Let  $\mathcal{H}(p, i)$ be the ID of the device that packet $p$ visited $i$ hops ago, for $0 \leq i < N_{history}$. When $i=0$, then $\mathcal{H}(p, 0)$ is  the device at which $p$ is currently located.  
    To make the context features relational, rather than use device IDs, we use a sequence of Boolean features, $b_i$, defined as:
    \begin{eqnarray}
      b_i = 
    \begin{cases}
        1, & \text{if } u = \mathcal{H}(p, i), \\
        0,              & \text{otherwise}. 
    \end{cases}
    \end{eqnarray}
    The use of packet history 
    reduces 
    unnecessary packet transmissions. For instance, even if a possible next hop device $u$ has promising features for reaching the destination, if packet $p$ recently visited $u$, then $u$ may be a less good next hop than it seems based solely on $u$'s other feature values.

\end{enumerate}
\vspace{0.1in}


{\em Basic features.} 
We designate a  subset of the above features as {\em basic features} (marked in Table \ref{tab:normalization}) that we use in all of the trained DRL agents, and designate the remaining features as {\em additional features} that we evaluate through an ablation study in Section~\ref{sec:para-choice-ablation}. 
Specifically, five features, the packet time-at-device, Euclidean distance, timer transitivity, last inter-meeting time, and last meeting duration features, are designated as additional features, and the remaining seven features are designated as basic features.
We separate out Euclidean distance as an additional  feature since $x$ and $y$ location coordinates may not always be available. In such situations, the timer transitivity and other timing related features (i.e., packet time-at-device, last inter-meeting time, and last meeting duration) could be used instead.

%

{\em Feature estimation.}
All features are  estimated using local exchange of information between neighboring devices. A device discovers its neighbors when another device enters or leaves its transmission range through the use of ``heartbeat" control messages.  Suppose there is a packet $p$ with destination $d$ at device $v$, and  suppose $v$ has neighbors $u \in Nbr(v)$. Device  $v$ obtains the following information from each neighbor $u$: i) the features  ${f}_{device}(u, d) \cup f_{path}(u, d)\cup f_{nbrhood}(u, d)$, 
ii) $u$'s current $x$ and $y$ location coordinates, timestamped with $u$'s current clock, and iii) the $x$ and $y$ location coordinates for every other device $w$,   which $u$ has either recorded directly from $w$ or received indirectly from another device, along with the recording's timestamp, i.e., 
the time on $w$'s clock of when the coordinates were recorded.
Device $v$ then uses ii) and iii) to update its recording of the
$x$ and $y$ location coordinates for every other device, overwriting older recordings with more recent recordings for a device $w$, comparing the  timestamps associated with the recordings. Because these timestamp comparisons always compare timestamps received from the same device $w$, no clock synchronization is needed. 


{\em Feature normalization}. 
Our goal when normalizing features is to re-scale them into the range of approximately 0 to 1. Mobility makes  normalization  challenging as the ranges of  the raw feature values, such as for node degree, may be very different in different mobile networks. To address this, we make the normalization a function of network properties when possible, such as the (approximate) number of devices in the network or the (approximate) size of the area in which devices are moving or the maximum expected inter-meeting time between pairs of devices. In this way, the normalization can better adapt to new network environments.
Table \ref{tab:normalization} summarizes how we normalize features  in our simulations.

During training vs. testing it can be useful to scale some features slightly differently.
For instance, during training, a packet's time-to-live (TTL) should be sufficiently long to still allow packets to be delivered to the destination despite some random exploration, but also short enough to allow packet drops due to expired TTLs. Conversely, during testing, dropping packets due to expired TTLs may not be desirable as that would skew the delay results for delivered packets as dropped packets cannot be usefully counted. 
During testing, we use a larger initial TTL than during training (see Table \ref{tab:parameters}),
but we re-scale the raw TTL before normalizing  to ensure that as long as it is within 
the TTL range used in training
then the TTL feature will have the same value as during training and zero otherwise.





\begin{table}[t]
\centering
\caption{For each feature $f_i$, normalization is done using $(f_i + 1) / (D +1)$, where $1$ is added to $f_i$ to avoid zero values for features and the value $D$ is given in the table. 
Neighborhood features are not normalized as they are a function of other normalized  features;  context features also are not normalized as they are Boolean valued. We also distinguish the {\em basic features} used by all DRL agents whose performance we evaluate in our simulations from the {\em additional features}, see Table \ref{tab:drl-strategies}.
}
  \begin{tabular}{lllc}
     \toprule
        {\bf Class}  &   {\bf Feature}         & {\bf $D$}    & {\bf Basic Feature?}  \\ 
    \midrule
       {\bf Packet}   & {Packet TTL}              & {300}     & {Yes}  \\  
       {\bf Packet}   & {Packet time-at-device}       & {200}       & {No}   \\ 
         {\bf Device} & {Queue length}            & {20}     & {Yes}\\
     {\bf Device}     & {Per-destination queue length} & {20}    & {Yes} \\
      {\bf Device}    & {$x$-coordinate location}  & {500m }     & {No}\\
     {\bf Device}     & {$y$-coordinate location}  & {500m }   & {No} \\
     {\bf Device}     &   {Node degree}            & {10}     & {Yes}\\
     {\bf Device}     &   {Node density}            & {$N$}    & {Yes} \\
       {\bf Path}     &   {Euclidean distance}       & { 2 }    & {No} \\
       {\bf Path}     &  {Timer transitivity}      & {800}      & {No}\\
 {\bf Path}     & {Last inter-meeting time}           & {800}     & {No} \\
 {\bf Path}     & {Last meeting duration}           & {100}     & {No} \\

     \bottomrule
  \end{tabular}
\label{tab:normalization}
\end{table}

\subsection{MDP Formulation}
\label{sec:mdp}

Let $Nbr(v)$ be the current neighbors of device $v$. Each individual packet agent $p$  currently located at device $v$ can choose between moving to one of $v$'s neighbors or staying at device $v$. Packet $p$'s actions therefore correspond to the set $Nbr(v) \cup \{v\}$.
We next define the states, actions, and reward function for our packet-centric DRL agent from  $p$'s point of view.

\begin{itemize}
\item {\em  States}. 
The state features that packet $p$ uses to make a decision are a function of features derived from packet $p$, $p$'s destination $d$, and $p$'s current device $v$: 
$f_{s} (v, p, d) = 
{f}_{packet}(p) \cup 
{f}_{device}(v, d) \cup 
f_{path}(v, d) \cup {f}_{nbrhood}(v, d)$. 

\item {\em  Actions}. 
The action features for each action $u$ in packet $p$'s action set $Nbr(v) \cup \{v\}$
 are defined by
$f_{a}  (u, p, d) =
{f}_{device}(u, d) \cup 
f_{path}(u, d) \cup
{f}_{nbrhood}(u, d) \cup
f_{context}(p, u)$.  
This action description 
reuses many of the same features as in the state description, but is defined in terms of a  device $u \in Nbr(v) \cup \{v\}$  rather than  just packet $p$'s current location at device $v$, and includes the context features.
\end{itemize}

\vspace{0.1in}
\noindent{\em Reward function.} Forwarding strategies for mobile wireless networks must trade-off 
competing goals for packet delivery, such as minimizing delivery delay while also minimizing resource usage like energy and link bandwidth. As a packet travels  to its destination, it must also assign credit or blame to the devices it passes through, depending on the success or failure of the packet to reach its destination.
To incorporate these forwarding goals and network resource
considerations into packet decision-making,  we design a {\em weighted reward function}. Using our reward function, more weight can be placed on higher priority forwarding considerations, such as to not waste resources by making unnecessary packet transmissions.

We define separate rewards for 
the action of a packet choosing to stay at its current device, $r_{stay}$, vs. moving to a neighbor device that is not the destination, $r_{transmit}$, as packet transmission requires expending energy. The ratio of $r_{stay}$ to $r_{transmit}$  determines the trade-off that the forwarding strategy learns between minimizing packet delivery delay vs. number of transmissions.  
For instance, by setting $r_{stay} = r_{transmit}$, the DRL agent would minimize delay, but ignore the number of transmissions made. 
For mobile networks, where forwarding loops and unnecessarily long paths can  easily arise, explicitly penalizing transmissions is important for learning a more efficient forwarding strategy. 
We define two other rewards, for actions that lead to a packet being delivered to its destination, $r_{delivery}$, or dropped, $r_{drop}  = r_{transmit}/(1-\gamma)$, where $\gamma \in [0,1]$ is the RL discount factor. The drop reward is  equivalent to receiving a reward of $r_{transmit}$ for infinite time steps. 
Our reward settings  are given in  Table \ref{tab:parameters}.

\vspace{0.1in}
\noindent{\em Actions vs. options.}
The sample estimate of expected return for an  option that starts at time step $t_i$ and ends at time step $t_j$ is (see \cite{sutton1999between}): 
\begin{align*}
y &=\sum_{k=t_i}^{t_j-1} \gamma^{k-t_i} \cdot r_{k} + \left [\gamma^{(t_j - t_i)} \cdot \max_{a' \in {\mathcal{A}(s_{t_j})}} Q(s_{t_j}, a') \right ]
\end{align*}
where $r_k$ is the reward at time step $k$ as defined above, and $s_{t_j}$ is the state encountered at time $t_j$, with $\mathcal{A}(s_{t_j})$ its actions. 
We use this $y$ as the output target for the neural network.

Options for packets begin and end at transmission, i.e., when a packet leaves a device. Options can consist of a packet being {\em delivered}, {\em dropped}, or {\em transmitted and then staying} for some number of time steps. 
There are two types of options in this domain: terminal (packet delivery or drop) and non-terminal (transitions from one device to another and then staying at the new device). 
On packet delivery or drop, the option takes only a single time step and the next state $s_{t_j}$ is the terminal state. 
The sample of return for delivery is 
$y = r_{delivery}$, and similarly for the drop option, $y=r_{drop}$.

Because transmit and then stay options have rewards that are constant for every time step over the life of the option after the first time step, all $r_{k} = r_{stay}$ in the option except at the first time step $t_i$, where $r_i = r_{transmit}$.
Therefore, on transitions where the packet does a transmission action and then stays for several time steps, the sample of return is:
\begin{align*}
y &= r_{transmit} + R_{stay}(t_j -(t_i+1)) +
\gamma^{(t_j-t_i)} \cdot \max_{a' \in {\mathcal A}(s_{t_j})} Q(s_{t_j}, a').
\end{align*}
where  $R_{stay}(t_j -(t_i+1))$ is the 
return from the stay actions taken over the course of the option, from time $t_i + 1$ to time $t_j$,
and is defined by
\begin{align*}
    R_{stay}(t_j -(t_i+1)) & = r_{stay} \cdot \frac{1-\gamma^{(t_j-(t_i+1))}}{1-\gamma}.
\end{align*}

Because reward is constant for all but the first timestep,  only the beginning and end of an option need be stored during training.
Every packet that remains in a device queue at the end of a training  round has an unfinished option. 
We remove such options from the data. 
As more data accumulates, including the end of the option, the newly finished options are used. 
From this point on, to be consistent with Sections \ref{sec:mobile-features} 
and \ref{sec:mdp}, we use ``actions" rather than ``options" to refer to extended-time actions.


\subsection{Decision-Making}
\label{sec:decision-making}
The DNN architecture we use to approximate the Q-value function for the DRL agent is shown in Fig. \ref{fig:routing-architecture}. 
Our DNN has four layers: input, expansion, compression, and output. 
Let $F= \mid f_s (\cdot)\mid + \mid f_a(\cdot) \mid$ be the number of input features and thus the size of the input layer. The expansion layer  has $10F$ neurons and the compression layer has $F/2$ neurons. 
The DNN then outputs a Q-value for each state, action pair, represented by the feature vectors $f_s(\cdot)$ and $f_a(\cdot)$.  
Each device has a copy of the same trained DNN that encodes the forwarding strategy, which allows decision-making to be done independently at each device.

Fig. \ref{fig:routing-architecture} also overviews how our DRL packet agent uses the trained DNN to make a forwarding decision.
To obtain the features $f_{s} ( \cdot)$ and $f_{a}  (\cdot)$ to input into the neural network in Fig.  \ref{fig:routing-architecture},
the DRL agent computes the following features  for packet $p$ with destination $d$ currently at device $v$:
${f}_{packet}(p)$,   ${f}_{device}(v, d)$,  and $f_{path}(v, d)$  using local information at device $v$;  ${f}_{nbrhood}(v, d)$, ${f}_{device}(u, d)$, $f_{path}(u, d)$, and ${f}_{nbrhood}(u, d)$ using the features received by $v$ from $u\in Nbr(v)$; and finally,
${f}_{context}( p, u)$ using only $u$'s  ID in addition to the information carried in packet $p$'s header fields.
The number of times the DNN is used to predict Q-values corresponds to the number of actions available to the packet. Using the DNN to separately make predictions for each action, rather than making predictions for all actions at once, allows the DRL agent to handle varying numbers of actions, and, correspondingly, varying numbers of neighbors (and therefore varying topologies).
During training, $\epsilon$-greedy action selection is used, with $\epsilon$ set as in Table \ref{tab:parameters}.

\subsection{Offline Training}
\label{sec:offline-training}
In a mobile wireless networks, devices can only exchange information (such as whether a particular packet reached its destination or was dropped)  using distributed communication. Consequently, it is typically not feasible to gather the information needed for training online due to limited network bandwidth. Instead, we use {\em offline training}.
This approach to training is particularly important as even a single mobile network scenario includes network connectivity changing over time and space, making it essential to be able to include such diversity during training.

During training, for each packet decision, regardless of the packet's device location, we record the state features, the action features for each action considered, the action selected, and the received reward in one data file. 
Doing so allows us to construct the state, action, reward, next state tuples, $(s, a, r, s')$, that are needed for training the DNN approximating the DRL forwarding strategy.
Because our relational features are device and network independent,  we are  able to train a single DRL agent using this file, via
experience replay. 
Every 1000 time steps (corresponding to one round), the DRL agent is trained using a newly initialized DNN, and trajectory samples are taken from the data file comprising all data so far collected. Samples are only taken for packets that have been dropped or delivered (and hence the full trajectory is contained in the data file) to allow the computation of the option returns which is a function of the time it takes for the packet to reach a terminal state. Then, this newly trained DNN is used by the DRL agent to make more decisions and collect more training data. In our training simulations in Section \ref{sec:simulation} we collect $T_{train} = 90,000$ time steps of data.

Once training has been completed,
the learned forwarding strategy is then copied to each device and used {\em independently} at each device by packets to choose next hops. We record forwarding performance statistics during training to determine which model to select. We currently focus on minimizing delivery delay and so select the model that has low delay during training while also making a good trade-off in terms of resources used due to packet transmissions. In Section \ref{sec:simulation}, we find this is generally around 60,000 time steps when looking at the training plots of Fig. \ref{fig:rl-train-results}.

\begin{table}[t]
\centering
\caption{{\small Simulation parameters.}}
\begin{tabular}{llc}
\toprule
{\em Symbol}       & {\em Meaning}                    & {\em Value} \\
\midrule
$N_{train}$     & \# of devices during training   & 25 \\
$N_{test}$     & \# of devices during  testing    & 25, 64, 100 \\
$X_{train}$         & Transmission range for training & 50m \\
$X_{test}$         & Transmission range for testing & 30m to  80m \\
$\epsilon_{train}$ & RL  exploration rate for training  & 0.1\\
$\epsilon_{test}$ & RL  exploration rate for testing  &  0\\
$\gamma$           & RL discount factor          & 0.99 \\
$TTL_{train}$      & TTL field initialization during training  & 300 \\
$TTL_{test}$      & TTL field initialization during testing  & 3000 \\
$r_{delivery}$     & RL delivery reward & 0  \\
$r_{stay}$         & RL stay reward   & -1 \\
$r_{transmit}$     & RL transmit reward   &  -1, -2, -10  \\
$r_{drop}$         & RL drop reward                & $r_{transmit} / (1-\gamma)$ \\
$N_{history}$      & Length of device visit history & 0, 5\\ 
$T_{train}$        & \# of time steps for training    & 90,000 \\
$T_{test}$        & \# of time steps for testing    & 100,000 \\
$T_{model}$        & \# of training time steps used by testing model   & 60,000 \\
$T_{cooldown}$     & \# of time steps at simulation end with no traffic            & 10,000 \\
$T_{round}$        & \# of time steps per round  & 1000 \\
-                  & DNN training dropout rate & 0.2 \\
\bottomrule
\end{tabular}
\label{tab:parameters}
\end{table}

\section{Simulation Results}
\label{sec:simulation}

In this section, we overview our simulation setup and describe our simulation results. 
Our goal 
is to evaluate the performance and generalization capabilities of our proposed DRL-based packet forwarding strategy in a mobile wireless network.
We first describe how we simulate a mobile wireless network (Section \ref{sec:methodology}). 
Then, we describe the different forwarding strategies whose performance we compare against our DRL agent (Section \ref{sec:forwarding-strategies}).
Next, we overview DRL training performance (Section \ref{sec:training-performance}).
Finally, we overview DRL testing performance as well as the performance of the other forwarding strategies (Section \ref{sec:testing-performance}).

\subsection{Methodology}
\label{sec:methodology}
Our simulations are done using a  custom discrete-time  packet-level network simulator that we have implemented in Python3. 
The DRL agents are trained  and all strategies are tested using this simulator. We use  Keras v.2.5.0 \cite{chollet2015keras} and Tensorflow v.2.50  \cite{tensorflow2015-whitepaper} to implement the DNN. Table \ref{tab:parameters} gives our simulation parameters.

\subsubsection{Device Mobility}
\label{sec:connectivity}

We use two widely-used mobility models: the steady-state {\em random waypoint (RWP)} mobility  model \cite{navidi2004stationary,navidi2004improving,le2005perfect} and the original {\em Gauss-Markov (GM)} mobility model~\cite{Liang99:Gauss-Markov}. In the RWP model, each device picks a random location and speed, and travels to the chosen location at the chosen speed. After that, a device pauses for a random  duration, and repeats the above process until the end of the simulation. 
In the GM model, each device is assigned to move with an initial speed and direction, which are then updated at fixed intervals. The updated speed and direction values are correlated with the previous values (see details in~\cite{Liang99:Gauss-Markov}). 


\begin{figure}[t]
\centerline{\hbox{
    \subfigure[RWP mobility model]
    {\includegraphics[width=1.15in]{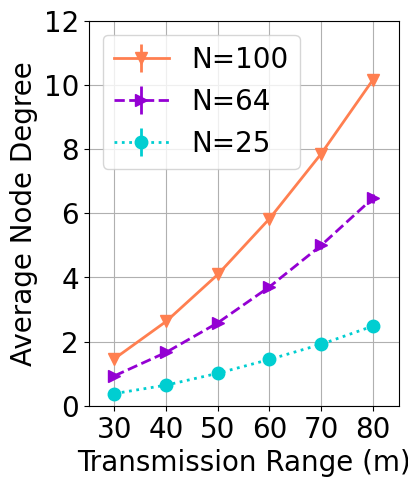}
    \hspace{-0.05in}
    \includegraphics[width=1.15in]{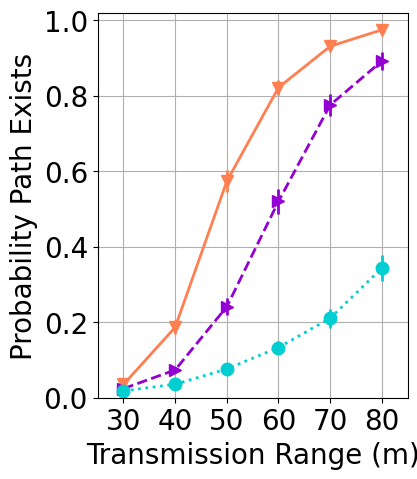}
    \hspace{-0.05in}
    \includegraphics[width=1.25in]{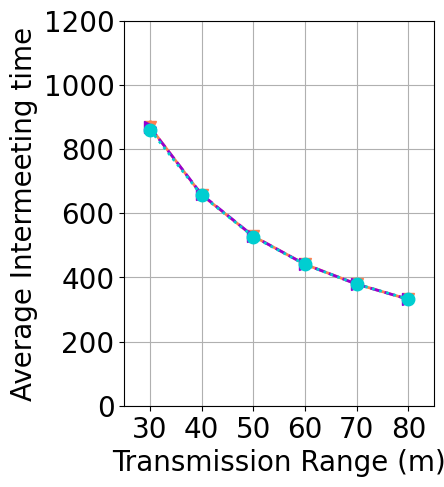}
    \hspace{-0.05in}
    \includegraphics[width=1.15in]{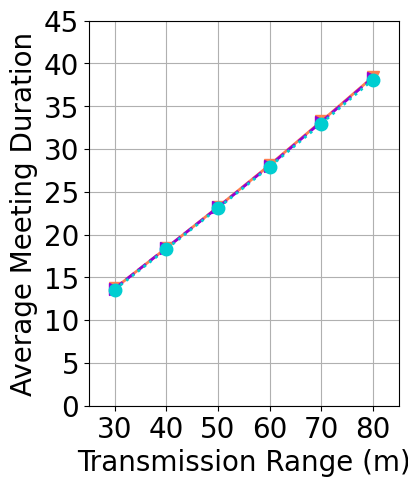}}
    }}
    \vspace{-0.05in}
    \centerline{\hbox{
    \subfigure[GM mobility model]
    {\includegraphics[width=1.15in]{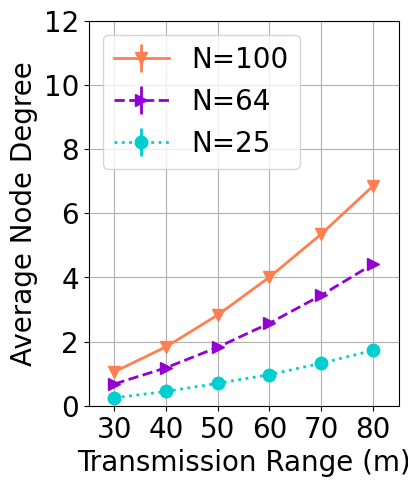}
    \hspace{-0.05in}
    \includegraphics[width=1.15in]{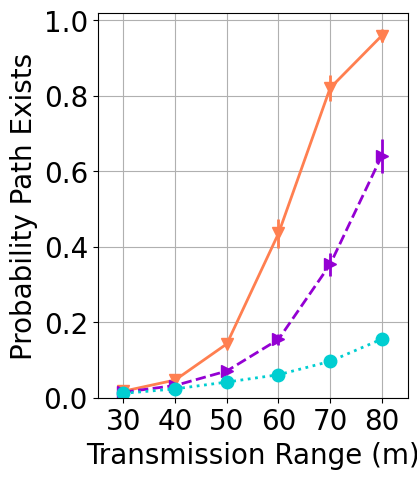}
    \hspace{-0.05in}
    \includegraphics[width=1.25in]{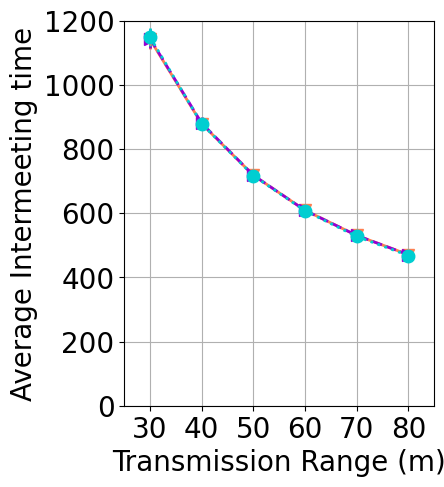}
    \hspace{-0.05in}
    \includegraphics[width=1.15in]{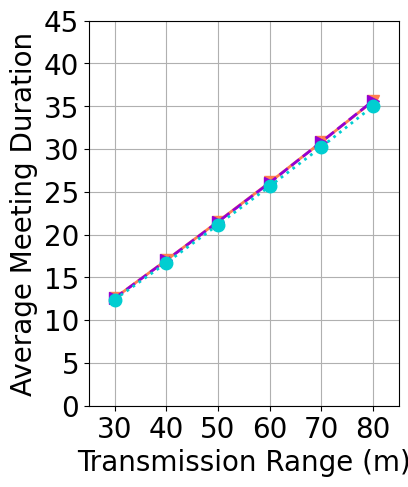}}
    }}
\caption{{The connectivity properties of the various network scenarios  considered in our simulations.
}}
\label{fig:rl-connectivity}
\end{figure}
In our simulations, we use BonnMotion \cite{aschenbruck2010bonnmotion} to generate 
mobility traces for devices moving under the above two models. Specifically, for both models, BonnMotion generates a mobility trace as a sequence of waypoints for each device. For example, suppose that device $v$'s mobility trace is $(t_0^v, x_0^v, y_0^v)$, $(t_1^v, x_1^v, y_1^v), \ldots, (t_n^v, x_n^v, y_n^v)$. Then device $v$ changes its waypoint at time $t_1^v, \ldots, t_n^v$, and moves from location $(x_i^v,y_u^v)$ to location $(x_{i+1}^v,y_{i+1}^v)$ at a constant speed during interval $[t_i^v, t_{i+1}^v]$.  
Since the time points at which the waypoint changes are decided by the mobility model, they  
 do not necessarily coincide with the per-second timesteps in our simulation. We therefore calculate the locations of each device $v$ at each timestep  $t$ by first finding the time interval $[t_k^v, t_{k+1}^v]$ in which $t$ lies in, and then calculating the location at time $t$ via a linear interpolation of $v$'s locations at $t_k^v$ and $t_{k+1}^v$.


Our simulations consider $N=25, 64$, and 100 devices moving in a 500m $\times$ 500m area. The transmission range is varied from 30m to 80m to obtain both poorly connected and very well connected mobile scenarios. For RWP traces,  devices move around at an average speed of 3 m/s with a speed delta of 2 m/s with no pause time.
For GM traces, devices move around at an average speed of 3.0 m/s with a  velocity standard deviation of 0.1 m/s.
Unlike the memoryless RWP model, the GM model uses a variable
$\alpha$ to control how much a device's past speed and direction are
considered when
setting the next speed and direction. 
We set $\alpha=0.6$ and update device
speed and direction every 30s in our simulations.

Fig. \ref{fig:rl-connectivity}(a) plots the average node degree, the probability that a path exists, the average inter-meeting time, and the average meeting duration for various numbers of devices $N$ and transmission ranges for the RWP model. 
The last two metrics are computed online between each possible pair of devices, and so are independent of the number of devices. As expected, the network  connectivity increases as the number of the devices $N$ and transmission range increase, leading to higher average node degree, higher probability of having a path between a pair of devices, shorter inter-meeting time, and longer meeting duration.  The corresponding quantities for the GM model are shown in Fig. \ref{fig:rl-connectivity}(b). 
We observe lower average node degree, lower probability of a path, longer inter-meeting time, and shorter meeting duration under the GM model than  under the  RWP model.     
This is consistent with the observation that for the same number of devices  and transmission range,  devices are more concentrated near the center of the simulation region under the RWP model than under the GM model~\cite{navidi2004stationary}. 

In our simulations, as listed in Table \ref{tab:normalization}, all training scenarios use $N_{train} = 25$ devices and a transmission range of $X_{train}=50$m. For testing, we vary the number of devices  $N_{test}$ from 25 to 100 and vary the transmission range $X_{test}$ from 30m to 80m. We generate separate mobility traces for the training and testing scenarios.


\subsubsection{Network Traffic}
\label{sec:traffic}
We vary the amount of traffic over time  by modeling flow arrivals, packet arrivals, and flow durations.  To generate traffic, we model flow arrivals using a Poisson distribution with parameter $\lambda_F = .001 N / 25$, scaling the number of flows (and thus the amount of congestion) as a function of the number of devices $N$ in the network. We model flow durations using an exponential distribution with parameter $\lambda_D=5000$ and packet arrivals on flows using a Poisson distribution with parameter $\lambda_P$ = 0.01. A simulation run starts with $\lambda_F \lambda_D$ initial flows. Each device has a queue with a maximum size of $B=200$ packets, beyond which additional packets are dropped. A packet's TTL field is initialized to $TTL_{train} = 300$ during training and $TTL_{test} = 3000$ during testing. We use a large testing TTL to ensure no packet is dropped due to an expired TTL. 

Because flow arrivals are a function of the number of devices in the network, 
not just network connectivity but also traffic congestion are varied in the different network scenarios used for testing. Thus, when evaluating generalization of a DRL agent trained on one scenario to other scenarios, we are evaluating not just generalization due to different connectivity and numbers of devices but also due to different amounts of traffic.

At each time step, every device in the network is given an opportunity to transmit up to $k=200$ packets in its queue. Given our network settings, this $k$ is sufficient for a device to transmit all of its packets,
avoiding the need for device decision-making. But if the number of packets in a device's queue were larger than $k$, the best $k$ packets could be forwarded where best is a function of the Q-value for each packet's best action  combined with a fairness measure to ensure each packet regularly gets a transmission opportunity. 
 


\begin{table}[t]
\centering
\caption{{DRL training scenarios. All training scenarios use $N_{train} =25$ devices, a transmission range of $X_{train}=$50m, and the {\em basic features}  specified in Table \ref{tab:normalization}. Scenarios are prefixed by type of mobility, either RWP or GM mobility. 
We use $T_{train}=60,000$ training time steps 
 for the trained DRL model based on looking at the training plots in  Fig. \ref{fig:rl-train-results}. The additional features used beyond the basic features are also listed for each scenario.}}
\label{tab:dynamics}
{\footnotesize
\begin{tabular}{lllll}
\toprule
Scenario            & $N_{history}$  & $r_{transmit}$ & Additional Features \\
\midrule
RWP-basic       & 5     & -2     & - \\
RWP-timer       & 5     & -2     & Timer   \\
RWP-timer$^+$      & 5     & -2      & Timer, Intermeeting, Duration, Time-at-device \\
RWP-dist    & 5     & -2         & Euclidean distance \\
RWP-dist-1    & 5     & -1        & Euclidean distance \\
RWP-dist-10   & 5     & -10       & Euclidean distance \\
RWP-dist-nohist    & 0     & -2        & Euclidean distance \\
RWP-dist-1-nohist   & 0     & -1        & Euclidean distance \\
RWP-dist-10-nohist    & 0     & -10        & Euclidean distance \\
GM-dist     & 5     & -2        & Euclidean distance \\
GM-dist-1     & 5     & -1           & Euclidean distance \\
GM-dist-10    & 5     & -10         & Euclidean distance \\
\bottomrule
\end{tabular}
}
\label{tab:drl-strategies}
\end{table}

\subsection{Forwarding Strategies}
\label{sec:forwarding-strategies}
In our simulations, we compare the performance of  five  forwarding strategies, including a delay minimizing strategy ({\em oracle}) and a transmission minimizing strategy ({\em direct transmission}) to give bounds on the performance of the DRL agent. We also compare with the utility and seek-and-focus strategies from \cite{singlecopy-spyro} as state-of-the-art strategies that trade-off delay with number of transmissions. 
Our goal is to understand which forwarding strategies have low packet delivery delay
while also making a good trade-off in terms of resources used due to packet transmissions.

\begin{figure*}[t]  
\centerline{\hbox{    
    \subfigure[RWP, packet delivery]
    {\includegraphics[width=1.5in]{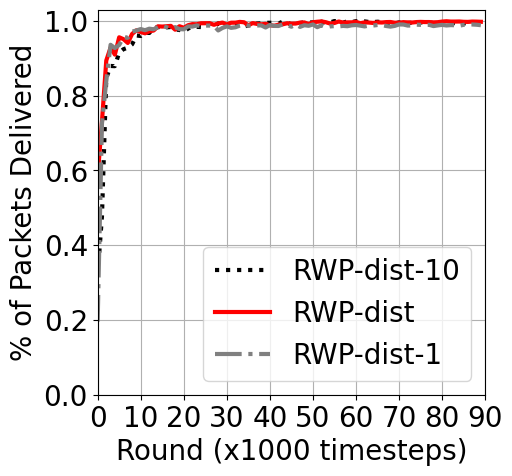}}
    \hspace{0.05in}    
    \subfigure[RWP, delay]
    {\includegraphics[width=1.5in]{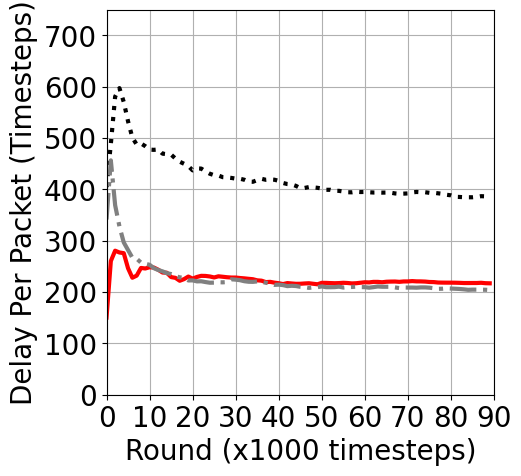}}
    \hspace{0.05in}        
     \subfigure[RWP, \# of forwards]
    {\includegraphics[width=1.5in]{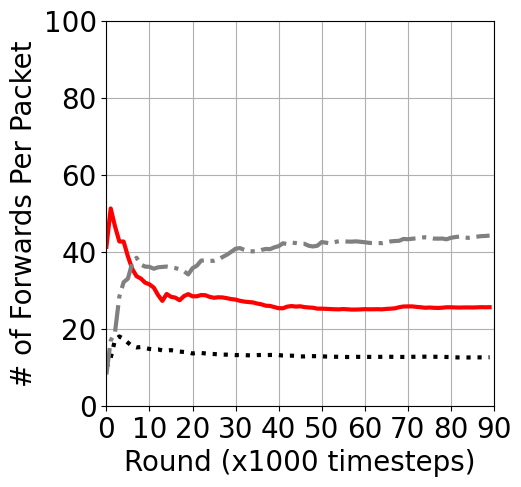}}}}
\centerline{\hbox{        
     \subfigure[GM, packet delivery]
    {\includegraphics[width=1.5in]{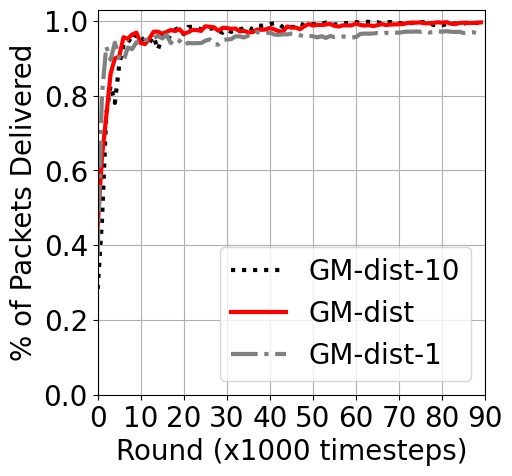}}
    \hspace{0.05in}    
     \subfigure[GM, delay]
    {\includegraphics[width=1.5in]{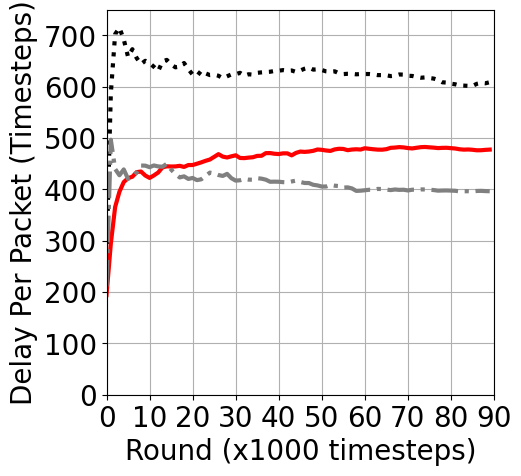}}
    \hspace{0.05in}    
     \subfigure[GM, \# of forwards]
    {\includegraphics[width=1.5in]{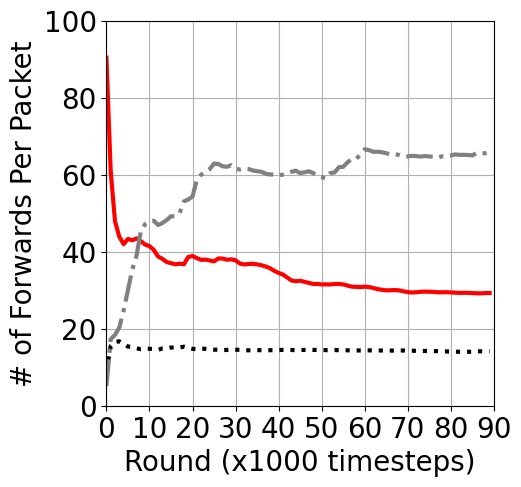}}}}
\caption{{DRL agent training performance. All training runs are done with $N_{train}=25$ and $X_{train}$=50m.}}
\label{fig:rl-train-results}
\end{figure*}

\vspace{0.1in}
\begin{enumerate}

\item{\em Oracle forwarding} uses  complete information about current and future network connectivity to calculate the minimal hop path that achieves the minimal delivery delay for each packet. To find these forwarding paths, we make use of epidemic routing~\cite{vahdat20:epidemic}. Epidemic routing creates many copies for a packet and distributes them to the network. For those packet copies that reach the destination with the minimum latency, we further find the packet copy that reached the destination with the minimum number of hops. Although the  oracle forwarding strategy minimizes delay, while maintaining a good trade-off in terms of network resources, it is not  practical to implement in real networks since it requires knowing the current and future network topology.

\vspace{0.05in}
\item {\em Direct transmission forwarding}  only forwards a  packet one hop, directly from  the source to the destination.
This is optimal when the goal is to minimize the number of  transmissions per packet.

\vspace{0.05in}
\item {\em Utility-based forwarding} \cite{singlecopy-spyro} 
maintains a timer at each device $v$ for every other device $d$ in the network, denoted as $\tau_v(d)$, which is the time elapsed since device $v$ last met device $d$.
We implemented the timer transitivity as defined in \cite{singlecopy-spyro}.
For many mobility models, a smaller timer value on average implies a shorter distance to the device, so the timer  evaluates the ``utility'' of a device in delivering a packet to another device. 
A device $v$ chooses a neighbor $u$ as the next hop for a packet if $u$ has the smallest timer to the packet's destination $d$ (among all neighbors of $v$) and $u$'s timer to $d$ is smaller than  $v$'s timer to $d$ by more than the utility threshold,  $U_{th}$. If no such neighbor exists, device $v$ does not forward the packet. 
We optimize $U_{th}$ for $N_{train}=25$ and $X_{train}=50$m, 
which are the same settings on which the DRL agent was trained.
The details of utility-based forwarding and the parameter values we use can be found in Appendix~\ref{appendix:util}. 



\vspace{0.05in}
\item{\em  Seek-and-focus forwarding} \cite{singlecopy-spyro}  combines the utility-based strategy with a random forwarding strategy. If the packet is perceived to be far from its destination, the packet is in the seek phase (random forwarding); 
otherwise, the packet is in the focus phase, and utility-based forwarding is used to choose  the next hop. Two additional parameters are used to prevent a packet from being stuck in a phase for a long time: $T_{focus}$ controls the maximum duration to stay in focus phase, and $T_{seek}$ controls the maximum duration to stay in re-seek phase before going to seek phase. 
All together, six parameters are used in seek-and-focus forwarding. 
Again, we optimize these parameters for $N_{train}=25$ and $X_{train}=50m$. 
Further details and the parameter values we use are given in Appendix~\ref{appendix:util}.


\vspace{0.05in}
\item {\em DRL forwarding} uses a  DRL agent to make forwarding decisions. We consider a number of different DRL agents trained on different scenarios, shown in Table \ref{tab:drl-strategies}. 
Specifically, we prefix these scenarios by the type of mobility, i.e., ``RWP'' or ``GM'', followed by the features used, e.g., basic features only (marked as ``basic''), or basic and additional features (marked as ``dist'' that use Euclidean distance as an additional feature, or ``timer'' that use timer transitivity feature, or ``timer$^+$'' that use timer transitivity feature and other timing information). By default, we use $N_{history}=5$ for context history; a scenario that does not use the context history feature is marked as ``nohist''. Last, a scenario that does not use the default value of $r_{transmit}=-2$ appends  a ``$-1$'' or ``$-10$'' to ``dist", representing the other two $r_{transmit}$ values of $-1$ and $-10$ that we explore.
During training, the DRL agent is essentially learning about different kinds of network connectivity over time, as well as different amounts of congestion over time (and the resulting variability in queue length).
In Fig. \ref{fig:rl-train-results}, we show DRL training performance, discussed further in the next section.

\end{enumerate}
\vspace{0.2in}

In our current implementation of the timer transitivity calculation, which is used by the utility-based and seek-and-focus forwarding strategies as well as by the DRL agent when using the timer feature, we assume the distance $d_{u,v}$ between two neighboring devices $u$ and $v$ is known. In practice, we can use the transmission range as an over-estimate of $d_{u,v}$. This is because timer transitivity is calculated only when  the two devices are within transmission range of each other, i.e., their distance from each other is no more than the transmission range. We expect that using transmission range as an approximation to distance will have only a negligible impact on the timer value, as 
the timer transitivity calculation itself is only a rough estimate since the true relationship between delay and distance in a given network depends on the actual movement of devices. Consequently, errors incurred in the approximation of $d_{u,v}$ from using the transmission range should have negligible impact.


\subsection{Training Performance}
\label{sec:training-performance}

Fig. \ref{fig:rl-train-results} overviews DRL training performance. Each training simulation is run for  $T_{train}$ time steps, where each time step corresponds to one second. Training is done every $T_{round}$ time steps using all data received up to that time step but using a randomly initialized DNN model. At each time step, the cumulative performance over all packets delivered up to the time step is shown.  Specifically, the performance metrics include the packet delivery rate, delay per packet delivered, and number of forwards per packet delivered. 

The top row of Fig. \ref{fig:rl-train-results} shows DRL training performance under the RWP model, for three scenarios, with the default amount of history that we use (i.e., $N_{history}=5$).  In addition, one of the scenarios uses the default $r_{transmit}=-2$, while the other two scenarios use $r_{transmit}=-1$ and $-10$, as marked in the legend. The bottom row of Fig. \ref{fig:rl-train-results} shows performance under the GM model for the same settings. For both mobility models, Fig. \ref{fig:rl-train-results} shows convergence of the learned DRL strategies for the different scenarios, with $r_{transmit}$ controlling the learned trade-offs between delay and number of forwards.
For a given reward setting, GM has a higher delay per packet delivered and a higher number of forwards per packet delivered than does RWP, which is consistent with the observation that GM leads to lower connectivity than RWP under the same setting (see Fig.~\ref{fig:rl-connectivity}).
We explore this resource-delay trade-off in our testing results in the next section.

For clarity, Fig. \ref{fig:rl-train-results} only presents the training performance for a subset of the DRL agents in Table \ref{tab:drl-strategies}. The training performance for the DRL agents not shown is similar to those in Fig. \ref{fig:rl-train-results} and we observed convergence for all of the DRL agents.

\subsection{Testing Performance}
\label{sec:testing-performance}

The focus of our testing simulations is to evaluate the performance and generalization capabilities of our proposed DRL approach.
All performance measures are computed over the delivered packets, such as the delay or number of forwards per packet delivered.
All performance metrics are averaged across all simulation runs for a given set of parameters, except for the maximum queue length which is the maximum seen in any of the simulation runs for a given set of parameters. 



\begin{figure*}  
\centerline{
    \hbox{    
    {\subfigure[$N_{test}=25$, $X_{test}=50$m]
    {\includegraphics[width=1.7in]{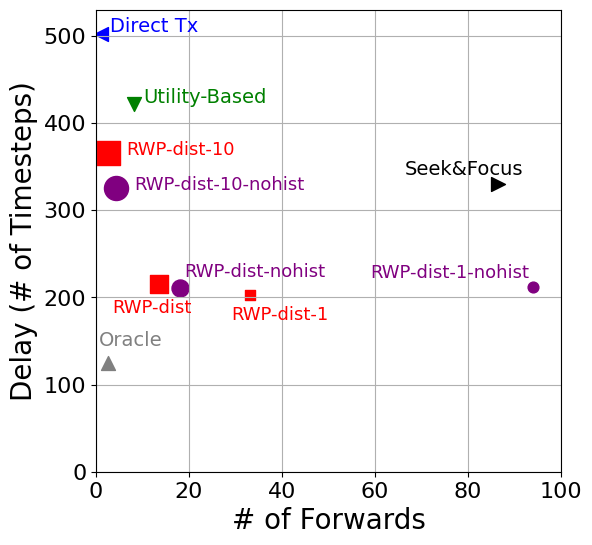}}}}
    \hspace{-0.17in}
    \hbox{    
    {\subfigure[$N_{test}=64$, $X_{test}=50$m]
     {\includegraphics[width=1.7in]{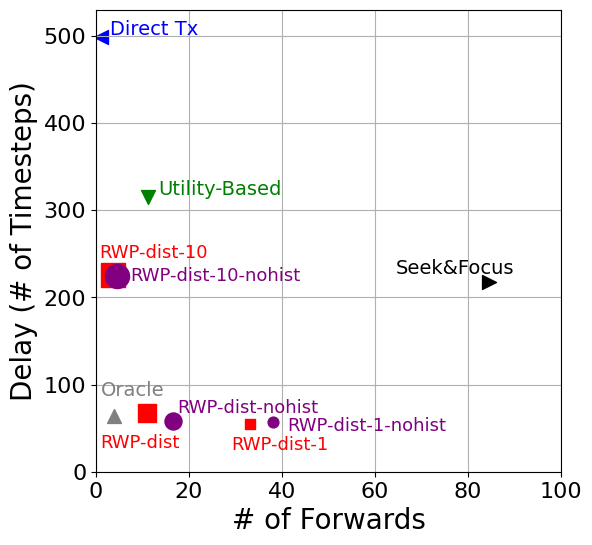}}}}
    \hspace{-0.17in}
    \hbox{    
    {\subfigure[$N_{test}=100$, $X_{test}=50$m]
     {\includegraphics[width=1.7in]{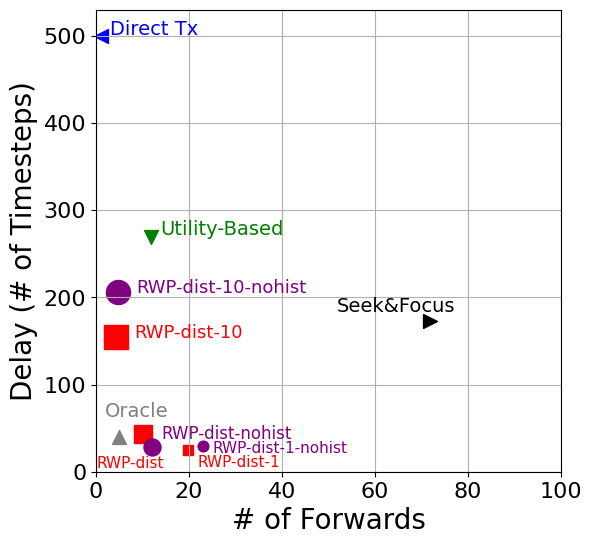}}}}   }
\caption{{Resource usage vs. latency using the RWP mobility model. Each  point is the average of 50 simulation runs. All DRL agents were trained with $N_{train}=25$ devices and $X_{train}$ = 50m and all packets were delivered and no packets were dropped.  Labels indicate the forwarding strategy.} 
} 
\label{fig:rl-test-reward}
\end{figure*}


\subsubsection{Impact of Parameter and Feature Choice} \label{sec:para-choice-ablation}

In this section, we investigate how the choice of training transmission reward, $r_{transmit}$, affects testing performance. We  also perform feature ablation to understand how different features impact performance.





\vspace{0.1in}
\noindent{\em Impact of transmission reward.}
Fig. \ref{fig:rl-test-reward} plots the trade-off between the delay (i.e., latency) and the number of packet transmissions (i.e., resource usage) for various forwarding strategies. For the DRL strategies, we consider multiple strategies that differ in their choice of $r_{transmit}$ and whether the context history feature is being used.  For comparison, we also show the trade-offs made by the transmission minimizing direct transmission strategy (``Direct Tx"), the delay minimizing oracle strategy, and the state-of-the-art utility-based and seek-and-focus strategies.
In the following, we consider the impact of the choice of $r_{transmit}$ on the delay vs. resource usage trade-off; the impact of the history feature is deferred to later. The performance comparison of DRL strategies with other strategies is discussed in detail in Section \ref{sec:res-RWP}.

Recall that the reward for staying at a device, $r_{stay}$, is fixed to $-1$. When $r_{transmit} = -1$, there is no penalty for transmission as a packet receives the same reward for transmitting as for staying at a node (i.e., $r_{transmit} = r_{stay}$).
Correspondingly, we see that for the DRL strategies using this reward setting, RWP-dist-1 and RWP-dist-1-nohist, 
delay per packet delivered is the lowest 
among the various DRL strategies, but  the number of forwards per packet is the highest. 
Conversely, for  $r_{transmit} = -10$, corresponding to RWP-dist-10 and RWP-dist-10-nohist,  there is a significant penalty for making a transmission, and so, while the delay per packet delivered is the highest for these DRL agents,  the number of forwards per packet is now the lowest. 
Henceforth, the  DRL agents used are trained with $r_{transmit} = -2$  as this provides a good trade-off between delay and number of transmissions. 


\vspace{0.1in}
\noindent{\em Impact of context history features.}
As shown in Fig. \ref{fig:rl-test-reward}(a), context history can also serve as a deterrent to unnecessary packet transmissions: the DRL agents that do not use context history (i.e., RWP-dist-nohist, RWP-dist-1-nohist, and RWP-dist-10-nohist) 
have a higher number of forwards, compared to the the DRL agents with $N_{history} = 5$ (i.e., RWP-dist, RWP-dist-1, and RWP-dist-10), for the same setting of $r_{transmit}$. This gap is most pronounced when there is no transmission reward penalty (i.e., when $r_{transmit} = -1$) and the network is sparse (i.e., for $N_{test} = 25$). While  $r_{transmit}$ penalizes packet transmissions, the
context history features themselves do not directly penalize transmissions. Instead, the context history features augment the state space to add context when actions are taken and ensure that actions that lead to unnecessary looping can be more easily identified. 

\begin{figure*}  
\centerline{\hbox{
    \subfigure[$N_{test}=25$, packet delivery]
    {\includegraphics[width=1.5in]{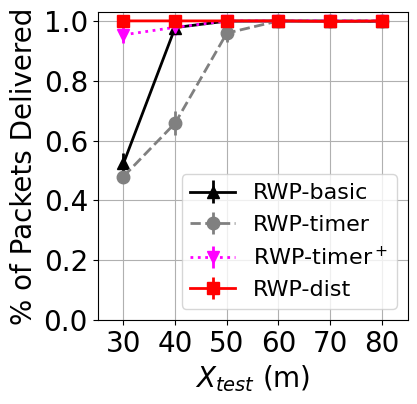}}
    \subfigure[$N_{test}=25$, delay]
    {\includegraphics[width=1.6in]{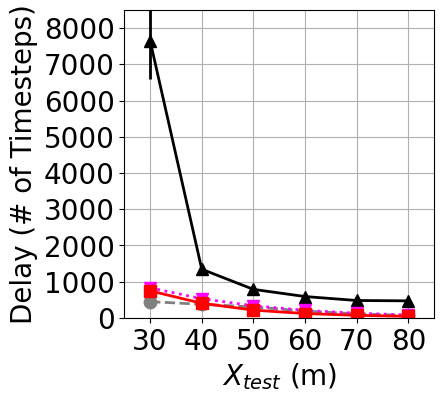}}
    \subfigure[$N_{test}=25$, \# of forwards]
    {\includegraphics[width=1.5in]{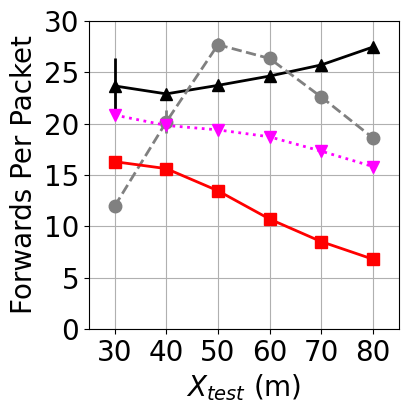}} }}
 \vspace{-0.1in}       
  \centerline{\hbox{
    \subfigure[$N_{test}=25$]
    {\includegraphics[width=1.5in]{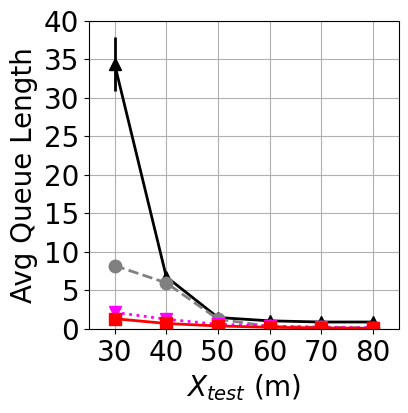}}
    \subfigure[$N_{test}=25$]
  {\includegraphics[width=1.55in]{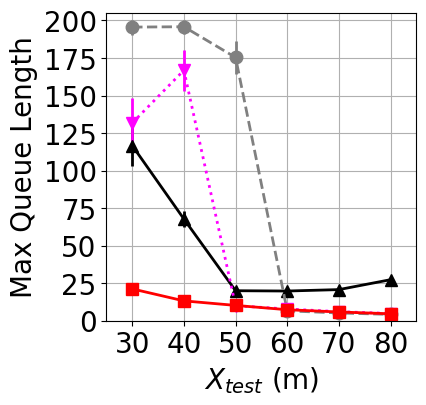}}    
    \subfigure[$N_{test}=25$, \# of drops]
    {\includegraphics[width=1.65in]{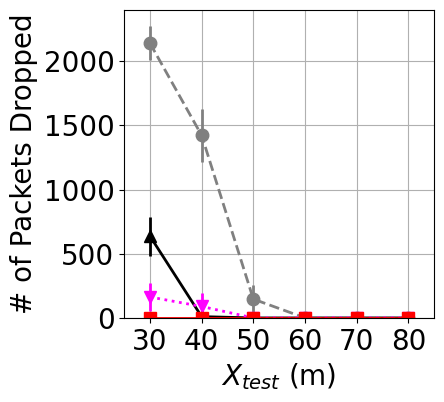}}}}
\caption{ 
Testing performance using the RWP mobility model, varying the features used by the DRL agents and the transmission range $X_{test}$ from 30m to 80m.  Each  point is the average of 50 simulation runs; 95\% confidence intervals are shown. All DRL agents were trained with $N_{train}=25$ devices and $X_{train}$ = 50m but with varying features, see Table \ref{tab:drl-strategies}. 
For the RWP-basic, RWP-timer, and RWP-timer$^+$ strategies, not all packets were able to be delivered by  the end of each simulation due to significant congestion and high delay. }
\label{fig:feature-ablation}
\end{figure*}

\vspace{0.1in}
\noindent{\em Impact of distance and timer features.}
Fig. \ref{fig:feature-ablation} looks at the impact of different features on performance. The RWP-basic strategy uses the fewest features, i.e., only the basic features listed in Table \ref{tab:drl-strategies}.  Fig. \ref{fig:feature-ablation}(b) shows that the 
RWP-basic strategy has larger average delay per packet delivered 
than any of the other strategies.  The RWP-timer strategy uses the timer transitivity feature in addition to the features used by the RWP-basic strategy. Fig. \ref{fig:feature-ablation} shows that adding the timer transitivity feature significantly reduces the delay compared to the RWP-basic strategy, but with increased forwards per packet delivered, and, more problematically, with significant numbers of packets dropped when the network topology is sparse (i.e., when $X_{test} < 60$m in Fig. \ref{fig:feature-ablation}(f)).

The RWP-timer$^+$ strategy 
uses the time-at-device, last inter-meeting time, and last meeting duration features in addition to the timer transitivity and other features used by the RWP-timer strategy.  
Fig. \ref{fig:feature-ablation} shows that these additional features give the RWP-timer$^+$ strategy a delay similar to that of  the RWP-timer strategy but with fewer forwards per packet and fewer  packet drops. 
In results not shown, we observe that for the RWP-timer$^+$ strategy, using the learned DRL model after 46,000 training time steps rather than 60,000 training results in no packet drops and stable queue lengths, though slightly 
larger delay but also significantly fewer forwards per packet delivered.
Consequently, there may be some overfitting with the RWP-timer$^+$  strategy  due to the 60,000 time steps of training. 

 Fig. \ref{fig:feature-ablation} shows that the RWP-dist strategy achieves the best performance for all of the performance metrics among all of the DRL strategies.
Comparing the performance of the RWP-timer$^+$  strategy with that of the RWP-dist strategy indicates that the timer transitivity feature can potentially be used as an  approximation to the Euclidean distance feature. In the rest of our testing results in the next sections, we focus on the DRL agents that use the distance feature in addition to the basic features.

\subsubsection{Results for RWP Mobility Model}  \label{sec:res-RWP}
In this section, we investigate
how well the learned DRL forwarding strategies generalize during testing using the RWP mobility model.

\begin{figure*}  
\centerline{\hbox{
    \subfigure[$N_{test}=25$]
    {\includegraphics[width=1.5in]{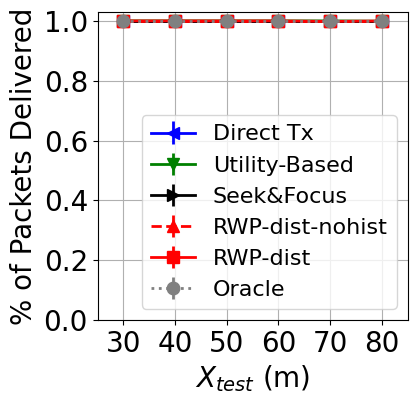}}
     \subfigure[$N_{test}=64$]
    {\includegraphics[width=1.5in]{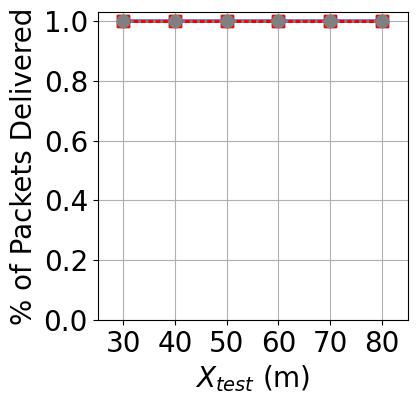}}
    \subfigure[$N_{test}=100$]
    {\includegraphics[width=1.5in]{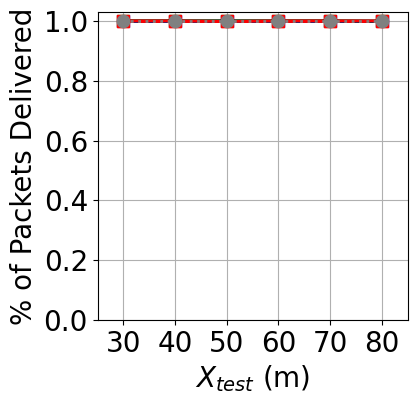}}  
}}
 \vspace{-0.1in}   
\centerline{\hbox{
    \subfigure[$N_{test}=25$]
    {\includegraphics[width=1.5in]{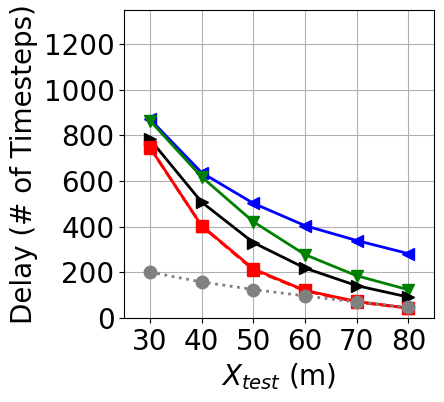}}
     \subfigure[$N_{test}=64$]
    {\includegraphics[width=1.5in]{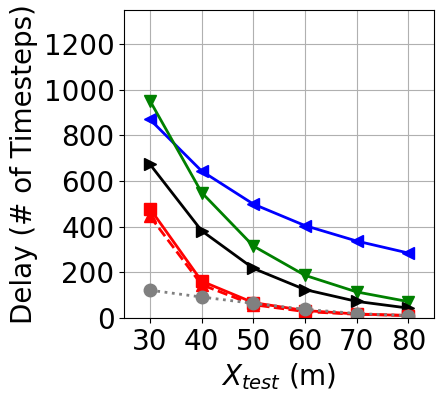}}
    \subfigure[$N_{test}=100$]
    {\includegraphics[width=1.5in]{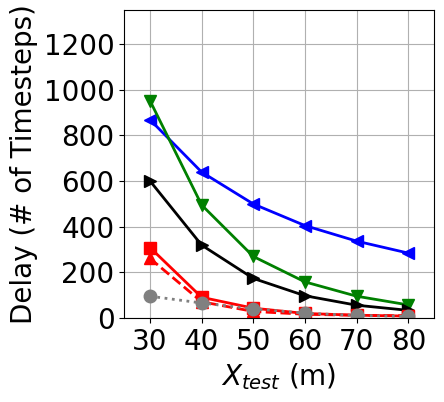}}  
}}
 \vspace{-0.1in}   
\centerline{\hbox{
  \subfigure[$N_{test}=25$]
    {\includegraphics[width=1.5in]{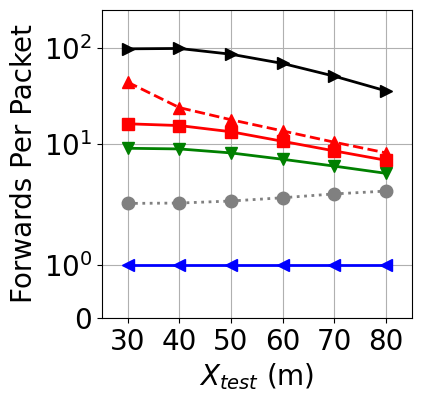}}
      \subfigure[$N_{test}=64$]
    {\includegraphics[width=1.5in]{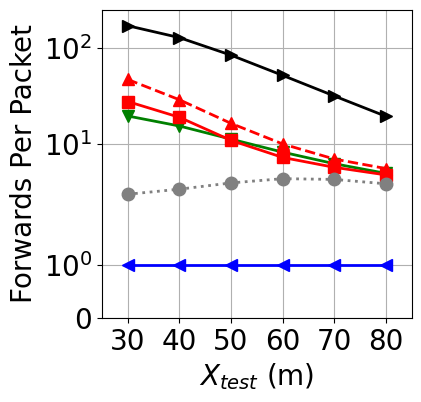}}
    \subfigure[$N_{test}=100$]
    {\includegraphics[width=1.5in]{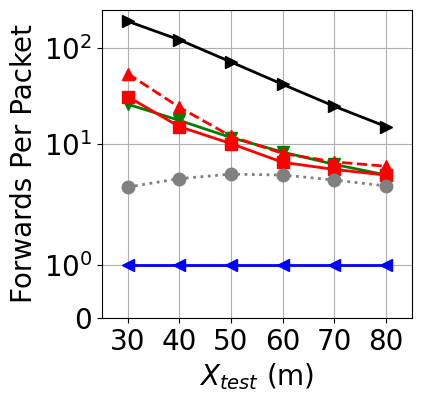}}
    }}
 \vspace{-0.1in}   
\centerline{\hbox{
   \subfigure[$N_{test}=25$]
    {\includegraphics[width=1.5in]{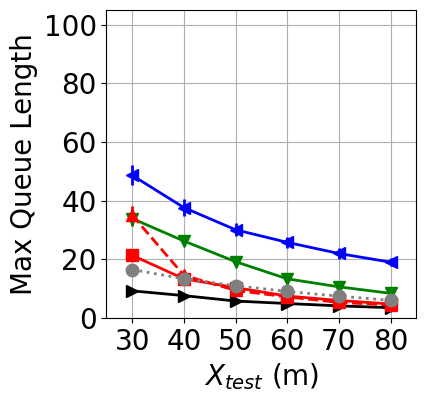}}
     \subfigure[$N_{test}=64$]
    {\includegraphics[width=1.5in]{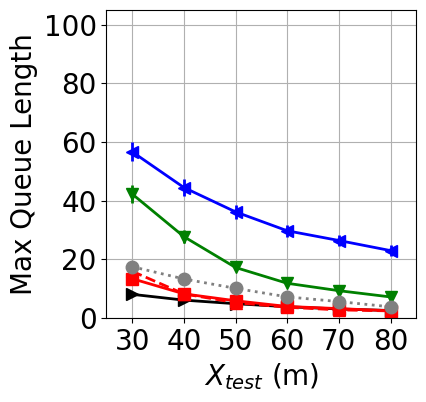}}
    \subfigure[$N_{test}=100$]
    {\includegraphics[width=1.5in]{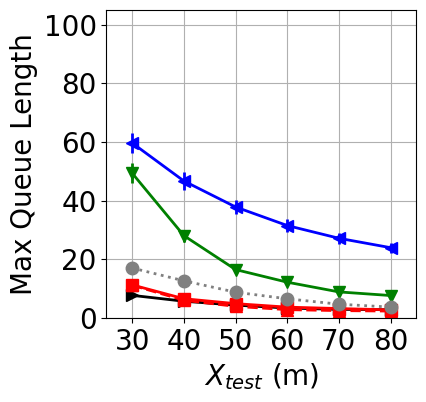}}
    }}    
\caption{{Testing performance using the RWP mobility model, varying the number of devices $N_{test}$ from 25 to 100 and the transmission range $X_{test}$ from 30m to 80m. Each  point is the average of 50 simulation runs; 95\% confidence intervals are shown. The legend shows DRL training conditions: both DRL agents were trained with $N_{train}=25$ devices and $X_{train}$ = 50m. All packets were delivered in these simulations and no packets were dropped.}
}
\vspace{-0.05in}
\label{fig:rl-test-results}
\end{figure*}

\vspace{0.1in}
\noindent{\em Overall results.} 
In Fig. \ref{fig:rl-test-results}, we plot the the testing performance of  two DRL agents, RWP-dist and RWP-dist-nohist, on the RWP model. 
The DRL agents were trained on the corresponding RWP scenarios described in Table \ref{tab:drl-strategies}, which use $N_{train}=25$ devices and a transmission range of $X_{train} = 50$m. 
The testing scenarios shown in  Fig. \ref{fig:rl-test-results} include 25, 64 or 100 devices (the largest network size we investigated) with transmission ranges varying from 30m to 80m, leading to various connectivity levels (see Fig. \ref{fig:rl-connectivity}).
Most of the testing scenarios differ from the training scenario in terms of the number of devices and/or the transmission range, as well as the number of traffic flows.  
We see in Fig. \ref{fig:rl-test-results} that our DRL agents are able to generalize  well from their training scenarios to the various testing scenarios, including those on which they were not trained. We also observe that our DRL agents often achieve packet delivery delays similar to the oracle strategy,  and outperform all other strategies in terms of delay. 

All forwarding strategies must make some kind of trade-off between packet delivery delay and number of packet transmissions. While the DRL agents use a reward function to navigate this trade-off, seek-and-focus uses a less straightforward approach. Among the six parameters of seek-and-focus, we observed that varying the forwarding probability $prob$ while fixing the other five parameters, see Appendix \ref{appendix:util}, provides one way to manage this trade-off. But there is no clear way to set 
these parameters in coordination with $prob$ to achieve a specific trade-off for a given network scenario.  One  solution  would be to  instead learn the optimal settings for the seek-and-focus parameters.


\vspace{0.1in}
\noindent{\em Impact of network connectivity.}
Fig. \ref{fig:rl-test-results} shows that as the network topology becomes more well connected (i.e., due to increasing number of devices $N_{test}$ or  increasing transmission range $X_{test}$), the DRL agents start to have delay per packet delivered similar to that of the oracle strategy, with not too many more forwards per packet delivered. As the network topology becomes more sparse (i.e., due to decreasing $N$ or decreasing transmission range  $X_{test}$), Fig. \ref{fig:rl-test-results} shows that the DRL agents start to have delay that is approaching the delay of the other strategies. The DRL agents have their highest delay per packet delivered for the $N=25$ and $X_{test}$ = 30m scenario which is the most disconnected testing scenario we consider (see Fig. \ref{fig:rl-connectivity}). Due to the few neighbors and relatively long inter-meeting times between pairs of devices for this scenario, including temporal neighborhood features would potentially be beneficial, as would be considering predicted future neighbors when choosing next hop actions.

Importantly, the DRL agents, trained on networks with $N_{train}=25$ devices, are able to generalize their learned forwarding strategies to networks with different numbers of devices ($N_{test}=64$ and $N_{test}=100$), as well as to the different numbers of actions that are available for packets at devices which depend on the number of neighbors that a device has.

\vspace{0.1in}
\noindent{\em Impact of network traffic.}
The number of flows (and therefore amount of traffic) increases as a function of the number of devices $N_{test}$ in the network (see Section \ref{sec:traffic}). Network congestion, however, increases as the transmission range $X_{test}$ decreases 
(due to correspondingly decreased network connectivity).
Fig. \ref{fig:rl-test-results} shows that the DRL strategies are able to generalize to  network congestion levels different from those on which they were trained.

\vspace{0.1in}
 \noindent{\em Impact of context history features.}
As we discussed earlier in Fig. \ref{fig:rl-test-reward}, 
we found the use of context history decreases the number of forwards per packet delivered. 
While the results we showed in Fig. \ref{fig:rl-test-reward}
are for a fixed transmission range of $X_{test}=50$m and varying numbers of devices $N_{test}$, 
Fig. \ref{fig:rl-test-results} validates this for other transmission ranges (for $X_{test}$ from 30m to 80m) as well as varying the number of devices.
Specifically, Fig. \ref{fig:rl-test-results}
shows that the DRL agent with no context history (i.e., RWP-dist-nohist) has a consistently higher number of forwards per packet delivered than the DRL agent with context history (i.e., RWP-dist) regardless of tranmission range and despite having a  similar delay  per packet. This is an indication that the use of context history improves generalization of the learned forwarding strategy. 

\vspace{0.1in}
\noindent{\em Queue stability.}
The stability of device queue lengths can be used to determine whether a network's capacity  can support a given traffic load and whether the forwarding strategy is appropriately managing this traffic. 
Figs. \ref{fig:rl-test-results} (j) to (l) show the maximum queue length seen in any of the simulation runs for a given set of parameters.  Though queue lengths increase as network connectivity decreases, we see that queue lengths are stable for all strategies, meaning that the forwarding strategy and network are able to support the traffic load. 
The direct transmission and utility-based strategies have noticeably larger queue lengths compared to the other strategies, with this gap increasing as the number of devices in the network increases. 

While not shown, average queue lengths are  small for all strategies, generally smaller than five packets. Note that the oracle strategy is implemented as a multi-copy scheme with a recovery latency of zero, i.e., when the first packet copy is delivered to the destination with minimum number of hops, all other copies are removed from the network. Consequently, the queue lengths for the oracle strategy include all packet copies present but only until the first  copy is delivered.

\begin{figure*}  
\centerline{\hbox{
    \subfigure[$N_{test}=25$]
    {\includegraphics[width=1.5in]{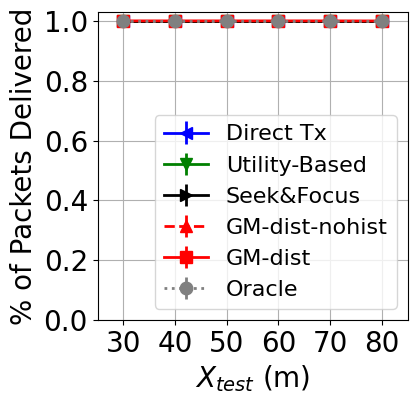}}
     \subfigure[$N_{test}=64$]
    {\includegraphics[width=1.5in]{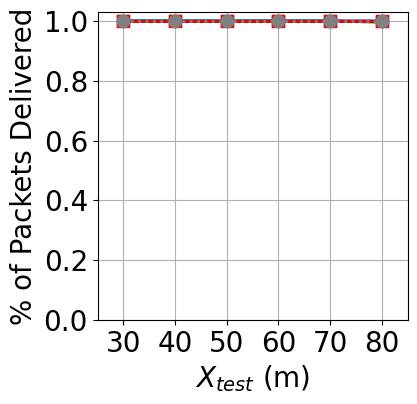}}
    \subfigure[$N_{test}=100$]
    {\includegraphics[width=1.5in]{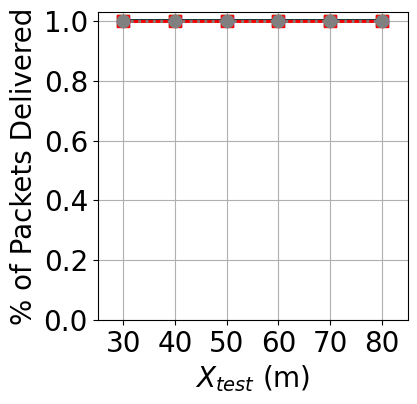}}  
}}
 \vspace{-0.1in}   
\centerline{\hbox{
    \subfigure[$N_{test}=25$]
    {\includegraphics[width=1.5in]{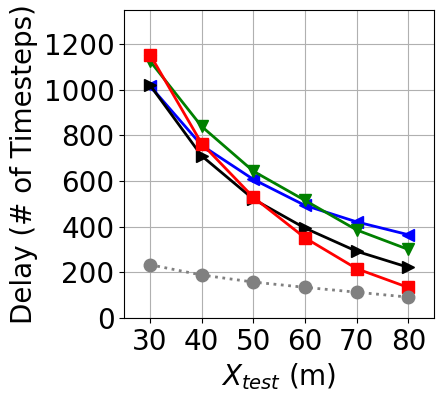}}
     \subfigure[$N_{test}=64$]
    {\includegraphics[width=1.5in]{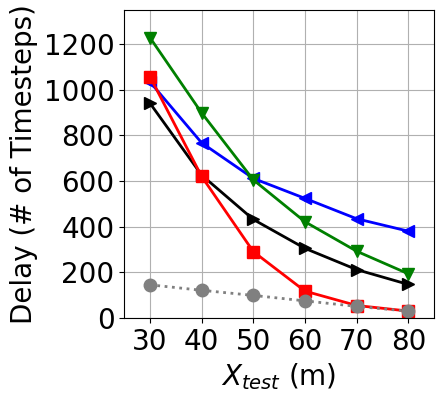}}
    \subfigure[$N_{test}=100$]
    {\includegraphics[width=1.5in]{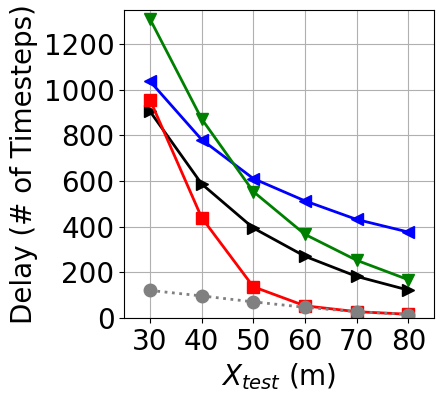}}  
}}
 \vspace{-0.1in}   
\centerline{\hbox{
  \subfigure[$N_{test}=25$]
    {\includegraphics[width=1.5in]{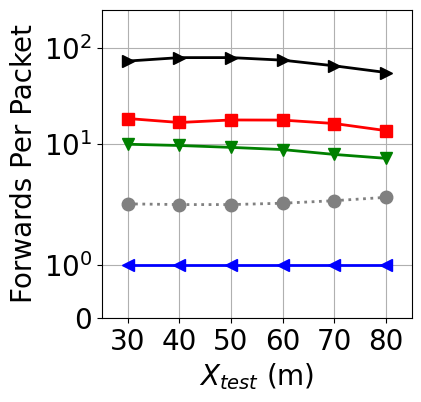}}
      \subfigure[$N_{test}=64$]
    {\includegraphics[width=1.5in]{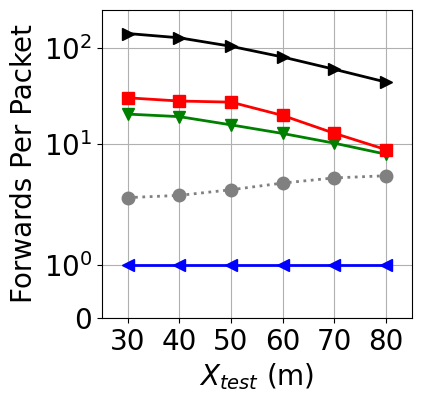}}
    \subfigure[$N_{test}=100$]
    {\includegraphics[width=1.5in]{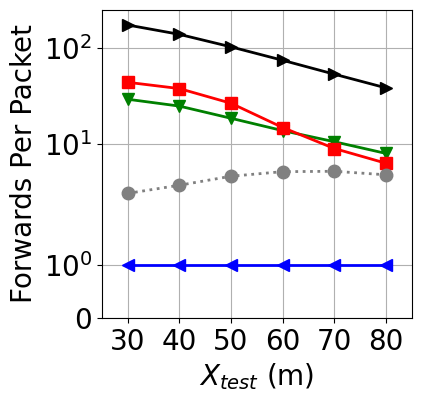}}
    }}
 \vspace{-0.1in}       
\centerline{\hbox{
   \subfigure[$N_{test}=25$]
    {\includegraphics[width=1.5in]{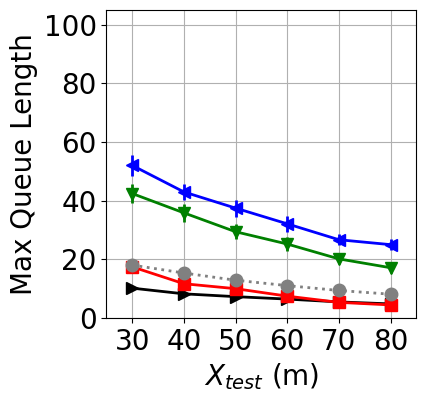}}
     \subfigure[$N_{test}=64$]
    {\includegraphics[width=1.5in]{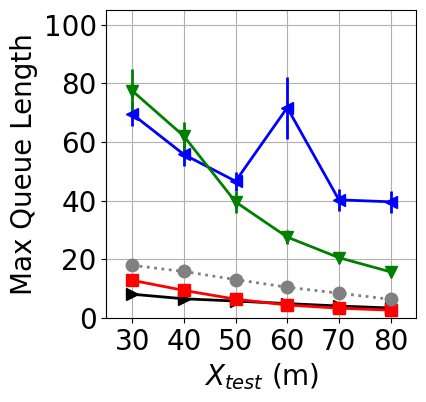}}
    \subfigure[$N_{test}=100$]
    {\includegraphics[width=1.5in]{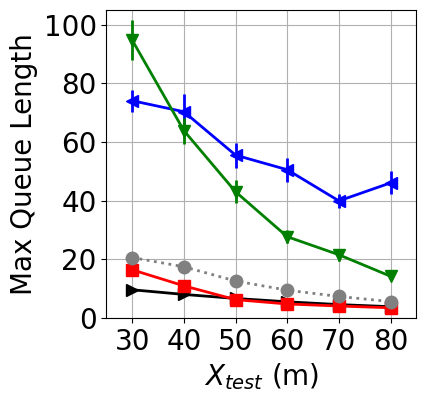}}
    }}      
\caption{{Testing performance using the GM mobility model, varying the number of devices $N_{test}$ from 25 to 100 and the transmission range $X_{test}$ from 30m to 80m. Each  point is the average of 50 simulation runs; 95\% confidence intervals are shown.  The legend shows DRL training conditions: the DRL agent was trained with $N_{train}=25$ devices and $X_{train}$ = 50m. All packets were delivered in these simulations and no packets were dropped.}}
\label{fig:rl-test-results-gm}
\end{figure*}

\subsubsection{Results for GM Mobility Model} 
In this section, we investigate how well our DRL forwarding strategy generalizes during testing using the GM mobility model. As shown in Fig. \ref{fig:rl-connectivity}, GM mobility gives rise to a significantly sparser network topology than does RWP mobility.

\vspace{0.1in}
\noindent{\em Overall results.} 
In Fig.  \ref{fig:rl-test-results-gm}, we plot the the testing performance of one DRL agent, GM-dist, on GM mobility.
Like with our RWP results, most of the testing scenarios differ from the training scenario in terms of the number of devices and/or the transmission range, as well as the number of traffic flows.  
Fig. \ref{fig:rl-test-results-gm} shows that our DRL agent is able to generalize  well from the training scenario to the various testing scenarios and often achieves packet delivery delays similar to the oracle strategy,  and outperforms all other strategies in terms of delay except for the very sparsest network scenarios.

\vspace{0.1in}
\noindent{\em Impact of network connectivity.} Like the RWP results in Fig. \ref{fig:rl-test-results}, the GM results in Fig. \ref{fig:rl-test-results-gm} show that as the network topology becomes more well connected, the DRL agent starts to have delay per packet delivered similar to that of the oracle strategy, with not too many more forwards per packet delivered. GM delay per packet delivered, however, is significantly higher than what is seen for the RWP results, due to the decreased connectivity of GM mobility.

\begin{figure*}  
\centerline{\hbox{
    \subfigure[$N_{test}=25$]
    {\includegraphics[width=1.23in]{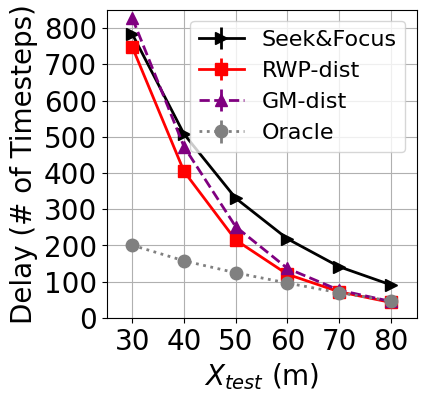}}
    \hspace{-0.1in}
    \subfigure[$N_{test}=25$]   
    {\includegraphics[width=1.22in]{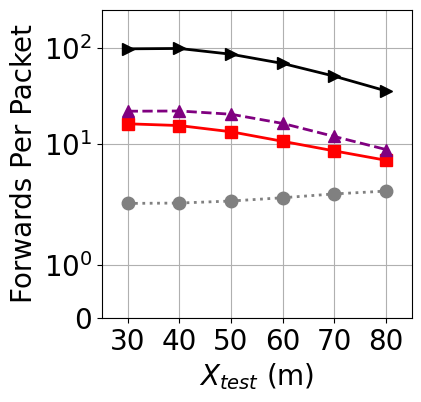}}
    \hspace{-0.1in}
  \subfigure[$N_{test}=25$]
    {\includegraphics[width=1.22in]{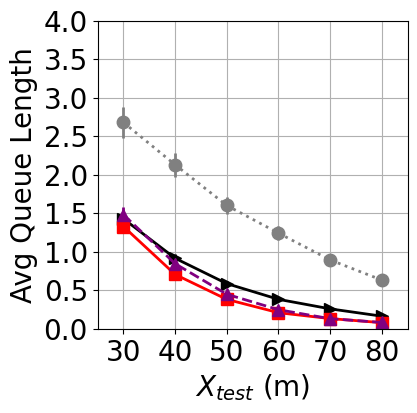}}
    \hspace{-0.1in}    
   \subfigure[$N_{test}=25$]
    {\includegraphics[width=1.2in]{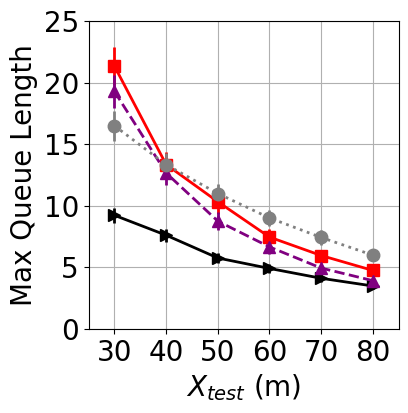}}
    }}    
 \vspace{-0.1in}
\centerline{\hbox{
    \subfigure[$N_{test}=64$]
    {\includegraphics[width=1.23in]{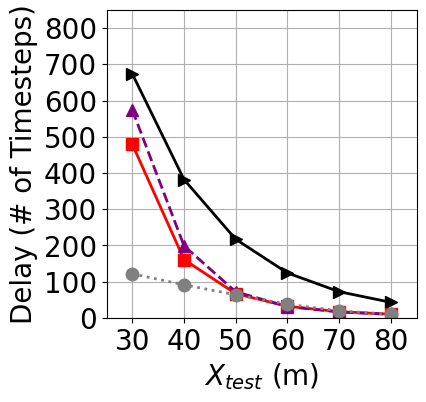}}
    \hspace{-0.1in}
    \subfigure[$N_{test}=64$]   
    {\includegraphics[width=1.22in]{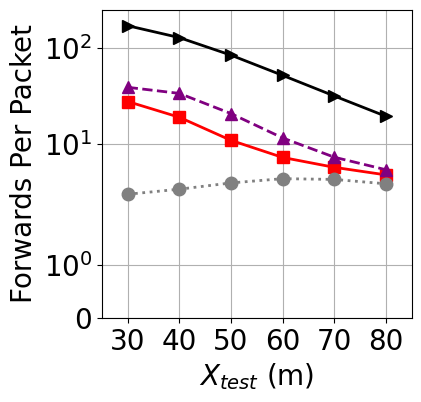}}
    \hspace{-0.1in}
  \subfigure[$N_{test}=64$]
    {\includegraphics[width=1.22in]{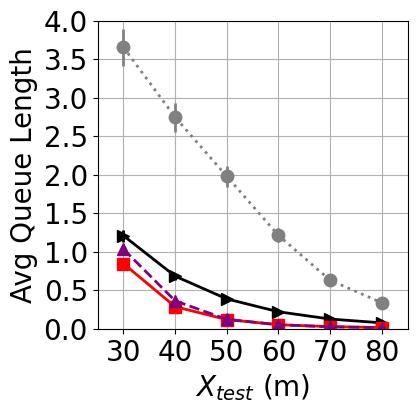}}
    \hspace{-0.1in}    
   \subfigure[$N_{test}=64$]
    {\includegraphics[width=1.2in]{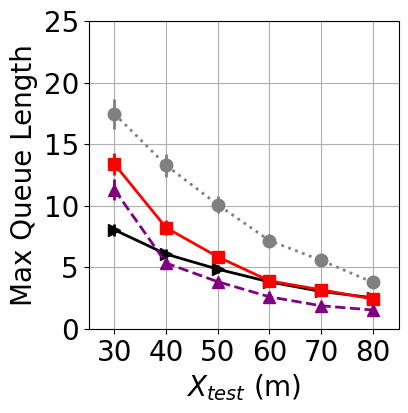}}
    }}        
\caption{
Cross-mobility testing performance: we apply GM-dist, a DRL strategy trained under the GM mobility model to  testing scenarios that use the RWP mobility model. For comparison, the results when using the RWP-dist strategy, i.e., a DRL strategy trained under the RWP mobility model (which matches the mobility model that is used in testing) are also shown.
Each  point is the average of 50 simulation runs; 95\% confidence intervals are shown. 
All packets were delivered in these simulations and no packets were dropped.}
\label{fig:rl-test-results-cross}
\end{figure*}

\vspace{0.1in}
\noindent{\em Impact of network traffic.}
Again, like the RWP results in Fig. \ref{fig:rl-test-results}, the GM results in Fig. \ref{fig:rl-test-results-gm} show that the DRL strategies are able to generalize to  different traffic and congestion levels, despite the sparser connectivity of the GM mobility model.

\vspace{0.1in}
\noindent{\em Queue stability.} 
While the queue lengths for the RWP results shown in Fig. \ref{fig:rl-test-results} are relatively stable, the GM results in Figs. \ref{fig:rl-test-results-gm} (j) to (l) show higher queue lengths and, in particular, noisier queue lengths for the direct transmission strategy. 
Average queue lengths, again not shown, are still generally less than 5 packets. 

\vspace{0.1in}
\noindent{\em Cross-mobility model generalization.} 
In Fig. \ref{fig:rl-test-results-cross}, we compare the testing performance of two DRL agents, GM-dist and RWP-dist, on the RWP mobility model. While the RWP-dist strategy was trained on the RWP model, the GM-dist strategy was trained on the GM model.
Our goal here is to confirm that our DRL-based approach to forwarding is able to generalize not just to different levels of connectivity and congestion for a given mobility model but also to different mobility models.
Indeed, we observe in Fig. \ref{fig:rl-test-results-cross} that the GM-dist strategy performs quite similarly to the RWP-dist strategy in terms of delay per packet delivered, with only slightly higher numbers of forwards. Interestingly, we also observe that the GM-dist strategy leads to slightly smaller maximum  queue lengths compared to that of the RWP-dist strategy, poentiatlly due to the GM-dist strategy training on the slightly sparser (and therefore more congested) GM mobility model.

\section{Conclusions and Future Work}
\label{sec:conclusions}
In this work, we have shown that it is possible to use DRL to learn a scalable and generalizable forwarding strategy for mobile wireless networks. We leverage three key ideas: i) packet agents, ii) relational features, and iii) a weighted reward function.
Our results show that our DRL agent generalizes well to scenarios on which it was not trained, often achieving delay similar to the oracle strategy and outperforming all other strategies in terms of delay including the state-of-the-art seek-and-focus strategy \cite{singlecopy-spyro}. 
The key ideas of our approach are generally applicable to other decision-making tasks in mobile wireless networks.


  
There are a number of research directions we would like to explore in future work.
We expect that as the kinds of mobility that the DRL agent sees become more diverse, more features will also be needed to characterize the key differences in mobility and enable the DRL agent to generalize to a wide variety of mobile networks. Additionally, targeted sampling of the feature space during training would be helpful to ensure that a diversity of network connectivity and congestion is seen.
We would also like to explore device decision-making to complement our packet agents. For instances, devices could learn which subsets of packets should get to make a forwarding decision when a transmission opportunity arises.
Finally, we would like to update our reward function to incorporate additional network considerations such as fairness.

\section*{Acknowledgements}
This material is based upon work supported by the National Science Foundation (NSF) under award \#2154190.
Results presented in this paper were obtained in part using CloudBank, which is partially supported by the NSF under award \#2154190. The authors also acknowledge the MIT SuperCloud and Lincoln Laboratory Supercomputing Center for providing HPC and consultation resources that have contributed to the research results reported within this paper.

\bibliography{routing,bing-ref,RLReferences}
\section*{Appendix}
\appendix

\section{Utility and Seek-and-Focus Strategies}
\label{appendix:util}

\begin{table}[t]
\centering
\caption{{\small Parameters settings.}}
{\footnotesize
\begin{tabular}{llc c}
\toprule
{\em Symbol}       & {\em Meaning}                         & {\em Utility} & {\em Seek\&Focus} \\
\midrule
$U_{th}$           & Utility threshold                     & 10 & 100 \\
$U_f$              & Focus threshold                       & - & 20 \\
$prob$             & Random forwarding probability         & - & 0.5 \\
$time\_until\_decoupling$ & Time before sending packet back to device  & - &10 \\
$T_{focus}$        & Max duration to stay in focus phase   & - & 10 \\
$T_{seek}$         & Max duration to stay in re-seek phase & - &50 \\
 
\bottomrule
\end{tabular}
}
\label{tab:param_settings-sf}
\end{table}

{\bf Utility-based forwarding} \cite{singlecopy-spyro} maintains a timer at each device $v$  for each device $d$ in the network, denoted as $\tau_v(d)$, which is the time elapsed since device $v$ last met device $d$.
We implemented the timer transitivity as defined in \cite{singlecopy-spyro}: when two devices, $u$ and $v$ encounter each other, if $\tau_u(d) < \tau_v(d) - t(d_{u,v})$, where $t(d_{u,v})$ is the expected time for a device to move a distance of $d_{u,v}$ (the distance between devices $u$ and $v$), then $\tau_v(d)$ is set to  
$\tau_v(d) = \tau_u(d) + t(d_{u,v})$.
The insight is that for many mobility models, a smaller timer value on average implies a smaller distance to the device, the timer  evaluates the ``utility'' of a device in delivering a packet to another device. 
A device $v$ chooses the next hop for a packet as follows: $v$ first determines which neighboring device, $u$, has the smallest timer to the packet's destination, $d$. If  $\tau_v(d) > \tau_u(d)+U_{th}$, i.e., the timer of $v$ to destination $d$ is larger than the timer of $u$ to $d$ by more than the utility threshold,  $U_{th}$, the packet is forwarded to device $u$; otherwise, the packet is not forwarded. 
We optimize $U_{th}$ for $N_{train}=25$ and $X_{train}=50$m,  
which are the same settings on which the DRL agent used in testing was trained; the parameter value we use  is shown in Table \ref{tab:param_settings-sf}.


{\bf  Seek-and-focus forwarding} \cite{singlecopy-spyro} combines the
utility-based strategy ({\em focus phase}) and random forwarding ({\em seek phase}). 
If the smallest timer (among all neighboring devices) to the destination is larger than the focus threshold $U_f$, the packet is in seek phase, forwarded to random neighbor with probability $prob$.
Otherwise, the packet is in the focus phase, and the carrier of the packet performs utility-based forwarding with utility threshold $U_{th}$.
In addition to  $U_f$,  $prob$, and $U_{th}$, Seek-and-focus  has three more parameters: the {\em time\_until\_decoupling} which controls the amount of time a device is not allowed to forward a packet back to a device it received the packet from, 
$T_{focus}$ which controls the maximum duration to stay in focus phase before going to re-seek phase (random forwarding of the packet to get out of a local minimum), and $T_{seek}$ which controls the maximum duration to stay in re-seek phase until going to seek phase (random forwarding of the packet until reaching a device with timer smaller than $U_f$). 
We optimize these parameters for $N_{train}=25$ and $X_{train}=50$m,  
which are the same settings on which the DRL agent used in testing was trained; the parameter values we use are shown in Table \ref{tab:param_settings-sf}.

\end{document}